\documentclass[sigplan,10pt]{acmart}

\copyrightyear{2025}
\acmYear{2025}
\setcopyright{rightsretained}
\acmConference[EuroSys '25]{Twentieth European Conference on Computer Systems}{March 30-April 3, 2025}{Rotterdam, Netherlands}
\acmBooktitle{Twentieth European Conference on Computer Systems (EuroSys '25), March 30-April 3, 2025, Rotterdam, Netherlands}
\acmDOI{10.1145/3689031.3696069}
\acmISBN{979-8-4007-1196-1/25/03}

\usepackage{enumitem}
\usepackage{amsthm}
\usepackage{subcaption}
\usepackage{pifont}

\usepackage{mathtools}
\usepackage{extarrows}

\usepackage{makecell}
\usepackage{multirow}
\usepackage{xcolor}
\usepackage{CJK}

\usepackage{color}
\usepackage{colortbl}

\usepackage{hyperref}

\newcommand{\para}[1]{\smallskip\noindent {\bf #1} }

\newcommand{\tool}{\textsc{Remix}}
\newcommand{\mixedSpec}{\textsc{mSpec}}

\newcommand{\zk}{ZooKeeper}
\newcommand{\tla}{TLA$^+$}
\newcommand{\bla}{\color{black}}

\newcommand{\blu}{\color{blue}}

\definecolor{Gray2}{gray}{0.75}
\definecolor{LightGray2}{gray}{0.85}
\definecolor{VeryLightGray2}{gray}{0.95}
\definecolor{blue-violet}{rgb}{0.54, 0.17,0.89}

\definecolor{bittersweet}{rgb}{1.0, 0.44,0.37}

\definecolor{chocolate}{rgb}{0.82, 0.41,0.12}

\definecolor{deepskyblue}{rgb}{0.0, 0.75, 1.0}

\newcounter{sectioncounter}[section]

\newcounter{globalcounter}

\newtheorem{theorem}{Theorem}

\theoremstyle{definition}
\newtheorem{defi}[theorem]{Definition}

\newtheoremstyle{named}{}{}{\itshape}{}{\bfseries}{.}{.5em}{\thmnote{#3 }#1}
\theoremstyle{named}
\newtheorem{namedtheorem}{Theorem}

\renewenvironment{proof}{{\noindent \it Proof. \\}\quad}{\newline\square\par}

\definecolor{midnightblue}{rgb}{0.1, 0.1, 0.44}
\newcommand{\hrefzkissue}[1]{\href{https://issues.apache.org/jira/browse/ZOOKEEPER-#1}{\textcolor{midnightblue}{ZK-#1}}}
\newcommand{\hrefzkpr}[1]{\href{https://github.com/apache/zookeeper/pull/#1}{\textcolor{midnightblue}{PR-#1}}}

\newcommand{\revision}[1]{#1}  
\providecommand{\revision}[1]{{{\color{blue} #1}}}
\newcommand{\revisionred}[1]{#1}  
\providecommand{\revisionred}[1]{{{\color{red} #1}}}

\begin{document}

\title{Multi-Grained Specifications for Distributed System Model Checking and Verification}

\author{Lingzhi Ouyang}
\orcid{0000-0001-7523-8759}
\affiliation{%
  \institution{SKL for Novel Soft. Tech.}
  \city{Nanjing University}
  \country{China}}
\email{lingzhi.ouyang@smail.nju.edu.cn}

\author{Xudong Sun}
\orcid{0009-0005-6734-0928}
\affiliation{%
  \institution{University of Illinois}
  \city{Urbana-Champaign}
  \state{IL}
  \country{USA}
}
\email{xudongs3@illinois.edu}

\author{Ruize Tang}
\orcid{0009-0001-0590-1620}
\affiliation{%
  \institution{SKL for Novel Soft. Tech.}
  \city{Nanjing University}
  \country{China}}
\email{tangruize@smail.nju.edu.cn}

\author{Yu Huang}
\orcid{0000-0001-8921-036X}
\affiliation{%
  \institution{SKL for Novel Soft. Tech.}
  \city{Nanjing University}
  \country{China}}
\email{yuhuang@nju.edu.cn}
\authornote{Corresponding author}

\author{Madhav Jivrajani}
\orcid{0009-0000-9170-6524}
\affiliation{%
  \institution{University of Illinois}
  \city{Urbana-Champaign}
  \state{IL}
  \country{USA}
}
\email{madhavj2@illinois.edu}

\author{Xiaoxing Ma}
\orcid{0000-0001-7970-1384}
\affiliation{%
  \institution{SKL for Novel Soft. Tech.}
  \city{Nanjing University}
  \country{China}}
\email{xxm@nju.edu.cn}

\author{Tianyin Xu}
\orcid{0000-0003-4443-8170}
\affiliation{%
  \institution{University of Illinois}
  \city{Urbana-Champaign}
  \state{IL}
  \country{USA}
}
\email{tyxu@illinois.edu}

\renewcommand{\shortauthors}{L. Ouyang, X. Sun, R. Tang, Y. Huang, M. Jivrajani, X. Ma, and T. Xu}

\begin{abstract}

This paper presents our experience specifying and 
    verifying the correctness of ZooKeeper,
    a complex and evolving 
    distributed coordination system.
We use \tla{} to model fine-grained behaviors of ZooKeeper
    and use the TLC model checker to verify its correctness properties;
we also check conformance between the model and code.
The fundamental challenge is to balance the granularity of specifications 
    and the scalability of model checking---fine-grained specifications
    lead to state-space explosion, while
    coarse-grained 
    specifications introduce model-code gaps.
To address this challenge, we write specifications with different 
    granularities for composable modules,
    and compose them into mixed-grained specifications
    based on specific scenarios. 
For example, to verify code changes,
    we compose fine-grained specifications of changed modules and 
    coarse-grained specifications that  
    abstract away details of unchanged code with preserved interactions.
We show that writing multi-grained specifications is a viable practice 
    and can cope with model-code gaps
    without untenable state space,
    especially for evolving software where changes are typically local and incremental.
We detected six severe bugs that violate 
    five types of invariants and verified their code fixes; 
    the fixes have been merged to ZooKeeper. 
We also improve the protocol design
    to make it easy to implement correctly.
\end{abstract}



\maketitle

\section{Introduction} 
\label{sec:intro}

Distributed systems that implement complex protocols 
    are notoriously difficult to develop and maintain.
It is non-trivial to implement non-deterministic, asynchronous behavior 
    and reason about correctness and fault tolerance.
Formal methods have been increasingly used
    to verify protocol designs~\cite{yao2024lvr,padon2020ivy,ma2019i4,ma2022sift,yao2022duoai,yao2021distai,hance2021finding,padon2016ivy} 
    and system code~\cite{hawblitzel2015ironfleet,wilcox2015verdi,sergey2018disel,honore2021ado,sharma2023grove}.
Recently, formal methods start to go 
    beyond academic research
    towards validating and verifying production distributed systems~\cite{bornholt2021lightweight,newcombe2015how_amazon,hackett2023understanding, kms_verification}.
For example, companies like Amazon, Azure, MongoDB, and LinkedIn 
    all use \tla{} to specify and model-check production systems~\cite{amazon-use-tlaplus,azure-use-tlaplus,mongodb-use-tlaplus,linkedin-use-tlaplus,oracle-use-tlaplus}.

This paper presents our experience specifying and 
    verifying ZooKeeper~\cite{hunt2010zookeeper},
    a complex, evolving 
    distributed coordination system which uses a totally ordered broadcast protocol, known as Zab~\cite{junqueira2011zab,reed08simple,junqueira2010dissecting}.
ZooKeeper is actively maintained as an open-source project~\cite{apache-zk};
    it is widely used in practice as a critical infrastructure system for storing 
    service metadata
    and for fault tolerance~\cite{chou2019taiji,hindman2011mesos,tang2015holistic,zookeeper-use-cases}.
We aim to verify not only the Zab protocol, but also the implementation in ZooKeeper.
We use \tla{} to specify ZooKeeper's behavior
    and use the TLC model checker to verify that 
    every execution satisfies all correctness properties~\cite{lamport2002specifying,tlc_tlaplus}.
To ensure that our model correctly describes the implementation's behavior,
    we run conformance checking to 
    detect discrepancies between the model and the implementation  (\S\ref{sec:conformance}).
Note that we target existing ZooKeeper code written in Java, 
    mostly maintained by developers with little formal method expertise;
    so, rewriting ZooKeeper with machine-checked proofs in verification-aware languages~\cite{leino2010dafny,lattuada2023verus,desai2013p} 
    or frameworks~\cite{honore2021ado, sharma2023grove}
    is not realistic in the short term.

The fundamental challenge is to balance the scalability of model checking
    and the granularity of specifications in modeling
    code-level implementation.
On one hand, fine-grained specifications, which model code-level behavior, 
    lead to {\it state-space explosion}. 
As a data point, model checking using TLC on ZooKeeper's official 
    system specification in \tla{}
    cannot finish in ten days 
    with a standard configuration of three nodes, three transactions, up to three 
    node crashes, and up to three network partitions~\cite{maric2017cutoff}.
In fact, this \tla{} specification omits 
    several important code behavior (\S\ref{sec:zk_sys_spec})
    like multithreading concurrency;
    modeling those would be even more costly.
Unfortunately, as subtle bugs often reside in deep states,
    extensive state-space exploration is inevitable.

On the other hand, coarse-grained specifications 
    introduce {\it model-code gaps}---the specification 
    does not effectively reflect 
    code-level implementation; 
    consequently, verification or
    model checking cannot capture subtle bugs whose manifestations are abstracted 
    away from the model.
We find that model-code gaps are prevalent.
One main reason is that implementations are highly optimized
    and the optimizations are rarely modeled in existing specifications.
A typical case is that an atomic action in the Zab protocol
    is implemented by several concurrent operations in ZooKeeper 
    (for performance optimization).
If the specification only models the protocol,
    bugs manifested via interleavings of concurrent operations
    cannot be exposed.
Such gaps are common in distributed systems projects:
    we inspected \tla{} specifications of MongoDB~\cite{mongodb-tlaplus}, CCF~\cite{microsoft-ccf-tlaplus},
    TiDB~\cite{tidb-tlaplus}, etcd~\cite{etcd-raft-tlaplus}, and CosmosDB~\cite{azure-tlaplus}; 
    local concurrency is often abstracted away. 

To address this challenge, we write {\it multi-grained specifications}, i.e.,
    multiple specifications with different 
    granularities for composable modules, 
    and compose them into {\it mixed-grained specifications}
    for specific scenarios.
For example, to verify a code change,
    we compose a mixed-grained specification using 
    fine-grained specifications of changed modules and 
    coarse-grained specifications that  
    abstract away details of unchanged code.
Essentially, this approach allows model checkers 
    to focus on the target modules with
    fine-grained modeling that reflects the implementation.
To enable multi-grained specifications,
    we write composable specifications for each module
    with an interaction-preserving principle,
    where a coarse-grained specification coarsens the corresponding fine-grained specification
    while preserving all actions whose effects are visible to the other modules.

To divide the specification into easily composable modules, we leverage an opportunity that
Zab, like other distributed protocols (e.g., Paxos, Raft, and 2PC), is designed to run in phases,
with clean boundaries between phases.
For example, Zab runs in four phases (Election, Discovery, Synchronization and Broadcast)
sequentially if no failure happens.
Thus, we decompose the ZooKeeper specification by phases and write multi-grained specifications
for each phase.

We show that multi-grained specification is a viable practice 
    and can effectively address model-code gaps
    without untenable state-space explosion,
    especially for evolving software where changes are typically local and incremental.
We are able to model low-level system behavior such as local concurrency
    in fine-grained specifications and use them to 
    create mixed-grained specifications with manageable state space.
This practice allows us to capture deep bugs that cannot be 
    found with \revision{existing} \tla{} specifications;
it also allows us to efficiently verify code changes and bug fixes, which 
    can introduce new bugs or fail to resolve the root cause.
The efforts of writing multiple specifications are manageable
    and are done incrementally.
As many distributed system projects have already adopted the practice of 
    writing \tla{} specifications, 
    we demonstrate the methodology to deepen \tla{} 
    specifications to verify system implementations.

We develop a framework for model checking and verification of distributed systems 
    with multi-grained specifications, named \tool{}.
It composes module specifications into mixed-grained specifications.
It also provides conformance checking to preclude deviations that could be introduced
    when writing new specifications.
Using \tool{}, we have detected six deep bugs in ZooKeeper code 
    and verified their fixes.
The effort also helps improve the Zab protocol
    to make it easy to implement correctly. 
Our evaluation shows that mixed-grained specifications can significantly 
    outperform \revision{existing} specifications 
    in verification effectiveness and efficiency.

This paper makes the following main contributions:
\begin{itemize}[leftmargin=*]
    \item We share our practice of writing multi-grained specifications
        with the interaction preserving principle;
    \item We present our mechanism and tooling for composing multi-grained specifications
        of different modules into the mixed-grained specification;
    \item We demonstrate the values of fine-grained modeling that reconciles 
        specifications and code implementation.
    \item \revision{We found six deep bugs in \zk{},
    verified their code fixes, and improved the protocol design.}
    \item \revision{Our artifact: \url{https://zenodo.org/records/13738672}.}
\end{itemize}

\section{Background} 
\label{sec:background}

\subsection{Existing Specifications}
\label{sec:existing_spec}

We started by writing \tla{} specifications 
    for the Zab protocol and the ZooKeeper system (its realization
    of the Zab protocol), respectively.
Both are now part of the official \tla{} specifications of the ZooKeeper projects.

\subsubsection{Protocol specification}
\label{sec:background:proto-spec}

\begin{figure}	
	\centering
    \begin{subfigure}[htbp]{0.495\textwidth}
		\includegraphics[width=\textwidth]{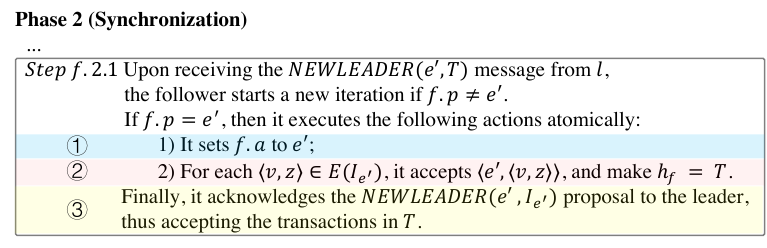}
		\caption{\bf Pen-and-paper description in the Zab paper~\cite{junqueira2011zab}}
        \label{subfig:zabpre-prot}		
        \vspace{10pt}
	\end{subfigure}
	\begin{subfigure}[htbp]{0.495\textwidth}
		\includegraphics[width=\textwidth]{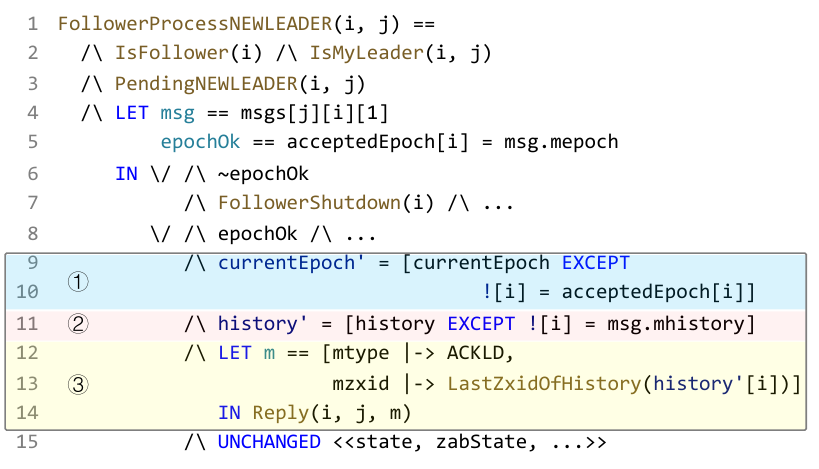}
		\caption{\bf Protocol specification in \tla{}}
        \vspace{-5pt}
        \label{subfig:zabpre-spec}
	\end{subfigure}
    \caption{\bf Pen-and-paper description and \tla{} specification of 
    Step $f.2.1$ in Phase 2 of the Zab protocol.}
    \vspace{2.5pt}
    \label{fig:proto-spec}
\end{figure}

The protocol specification follows the pen-and-paper description of the original Zab paper~\cite{junqueira2011zab}.
The goal is to formally describe and model check the protocol.
Figure~\ref{fig:proto-spec} shows the snippet of the protocol specification
    of Step $f.2.1$ in Phase 2 (Synchronization) of Zab~\cite{junqueira2011zab},
    where the follower is supposed to {\it atomically} execute two actions upon receiving a \texttt{\small NEWLEADER} message
    from the leader: 
    \ding{172} updating its current epoch 
    and \ding{173} accepting the leader's complete history.
In the protocol specification, we write the actions, together with the final acknowledgment,
    in a \tla{} atomic action \texttt{\small FollowerProcessNEWLEADER}.

Our \tla{} specifications also specify several missing components that are not described
    in the Zab protocol, e.g.,
    the Zab protocol does not describe leader election (it uses an assumed leader oracle)
    and does not describe certain unexpected cases (e.g., when the leader does not receive 
    sufficient acknowledgments from the followers).

\subsubsection{System specification} 
\label{sec:zk_sys_spec}
With the protocol specification, we 
    then wrote the system specification of ZooKeeper,
    as a precise, testable system design document~\cite{newcombe2015how_amazon}.
The system specification is developed based on the ZooKeeper source code,
    instead of the Zab paper.
For example, ZooKeeper implements a fast leader election algorithm, which is specified
    in the system specification,
    which refines the leader oracle in the protocol specification.

\begin{figure}	
	\centering
    \begin{subfigure}[htbp]{0.495\textwidth}
		\includegraphics[width=\textwidth]{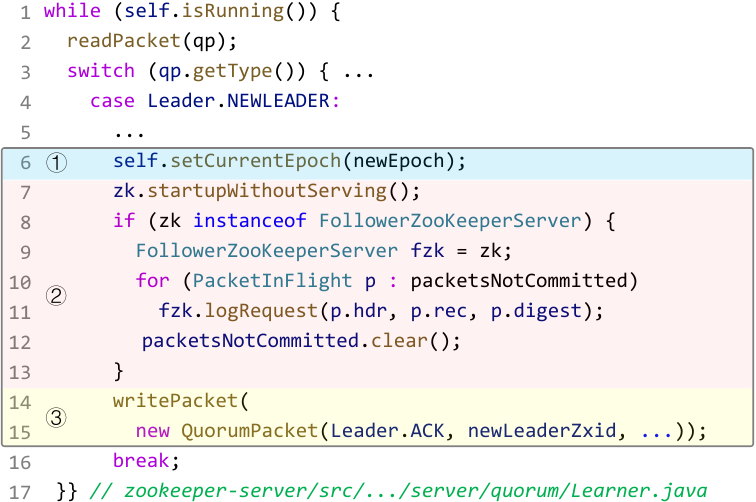}
		\caption{\bf Code implementation (v3.9.1) in Java~\cite{zookeeper-learner-code}} 
        \label{subfig:zab1.0-code}
	\end{subfigure}
	\begin{subfigure}[htbp]{0.495\textwidth}
		\includegraphics[width=\textwidth]{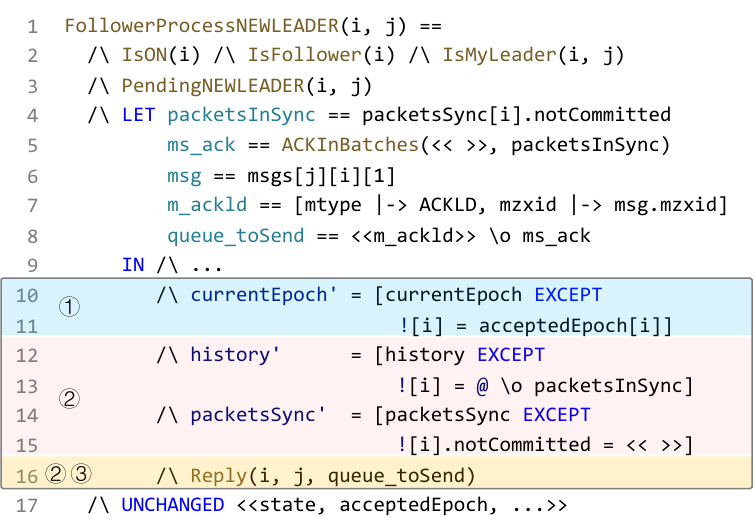}
		\caption{\bf System specification in \tla{}}
        \vspace{-5pt}
        \label{subfig:zab1.0-spec}		
	\end{subfigure}
	\caption{\bf Code implementation in Java and the corresponding system specification in \tla{} of 
        Step $f.2.1$ in Phase 2 of the Zab protocol in ZooKeeper.}
    \vspace{-5pt}
    \label{F:zab-example}
\end{figure}

Figure~\ref{subfig:zab1.0-code} shows the code snippet of ZooKeeper's implementation
    of Step $f.2.1$ of the Zab protocol.
When a follower receives the \texttt{\small NEWLEADER} message,
    it \ding{172} updates the current epoch, 
    \ding{173} logs every packet that is not committed,
    and \ding{174} replies to the leader.    
Figure~\ref{subfig:zab1.0-spec} shows the corresponding
    system specification.
Note that the system specification does not strictly refine the protocol specification.
The system implementation optimizes the synchronization using \texttt{\small NEWLEADER} as a signaling message
    without carrying concrete history, and the leader's history will be 
    synchronized in one of three modes (\texttt{\small DIFF}, \texttt{\small TRUNC}, and \texttt{\small SNAP}),
    depending on the differences between the history and the follower's latest transaction ID (\texttt{\small zxid}).
In this paper, since our goal is to verify the ZooKeeper system implementation,
    we start with the system specification instead of the protocol specification.

\subsection{Model-Code Gaps}
\label{sec:background:gaps}

Despite that the system specification effectively describes
    the ZooKeeper system,
    it still omits certain implementation details.
We present three common patterns of model-code gaps
    which in our experience is important to consider
    as they often induce tricky and error-prone 
    implementation.
Overlooking them would allow bugs to escape from model checking 
    and reduce the confidence of verification results.

\subsubsection{Atomicity} 
As a common model-code gap pattern, 
    an atomic action in the system specification is not guaranteed to 
    be atomically executed at the code level.
In this case, if a specified atomic action is partially executed and then interrupted,
    the intermediate states would be missed in model checking.
In Figure~\ref{subfig:zab1.0-spec}, \texttt{\small FollowerProcessNEWLEADER} is an atomic action:
    the state transitions from \ding{172} to \ding{173} to \ding{174} 
    are always atomically done in the system specification.
However, ZooKeeper's code-level execution does not guarantee such atomicity (Figure~\ref{subfig:zab1.0-code}).
We will show in \S\ref{sec:eval} that model-code gaps due to false atomicity
    would miss critical bugs.

We find that atomicity-related model-code gaps are common in \tla{} specifications
    of many distributed systems.
One reason is that every action in \tla{} is atomic
    so it is convenient to express logically connected steps 
    in an action.

\subsubsection{Concurrency}
We find that specifications commonly focus on 
    non-deterministic interleavings of actions {\it among} nodes, aka distributed concurrency~\cite{leesatapornwongsa2016taxdc}. 
\revision{Few model} local concurrency within a node (e.g., due to multithreading).
Instead, the specification often models locally concurrent events 
    in a single action with
    deterministic orders. 
However, if multithreaded code has non-deterministic behavior,
    model checking using such specifications would fail to explore bug-triggering states.
For example, in Figure~\ref{subfig:zab1.0-code},
    the follower's \texttt{\small QuorumPeer} thread calls \texttt{\small logRequest()} (line 11).
The implementation of \texttt{\small logRequest} 
    sends a logging request, 
    which would be asynchronously handled by a different thread. 
However, the above procedure is simplified 
    as the state transition of appending all uncommitted requests to the follower's history 
    (line 12-13 in Figure~\ref{subfig:zab1.0-spec}), 
    with sending replies to the leader in a deterministic order (line 16 in Figure~\ref{subfig:zab1.0-spec}).
Consequently, checking the specification will miss many possible states of asynchronous logging (\S\ref{sec:eval}).

Similar to atomicity, we find that model-code gaps related to local concurrency
    are common in \revision{existing} \tla{} specifications of many other distributed systems projects.

\subsubsection{Missing state transitions}
State transitions in the specification may be overly simplified compared to the code implementation.
For example, the follower in \zk{} would reply \texttt{\small ACK} upon receiving an \texttt{\small UPTODATE} message;
in the specification, the follower does not reply \texttt{\small ACK} to \texttt{\small UPTODATE} for simplicity.
Missing state transitions can cause model checking to miss possible states,
    \revision{and} meanwhile, explore false states that cannot be reached by code-level executions.

\subsection{Challenges}

The prevalence of model-code gaps in existing specifications 
    indicates the need to further 
    model important, fine-grained behavior like non-atomic updates and concurrency 
    for verifying code implementation
    (which is uncommon in \revision{existing} \tla{} specifications).
However, doing so would significantly increase state space,
    resulting in state-space explosion.
Currently, using TLC to model check the system specification 
    of ZooKeeper (\S\ref{sec:zk_sys_spec})
    cannot finish in ten days 
    with a standard configuration (three nodes, three transactions, up to three 
    node crashes, and up to three network partitions)~\cite{maric2017cutoff}.
How to balance the granularity of the specification
    and the scalability of model checking 
    is a key challenge.

\section{Writing Multi-Grained Specifications}
\label{sec:spec}

We write {\it multi-grained specifications}, i.e., multiple specifications with different 
    granularities for composable modules, 
    which can be composed into {\it mixed-grained specifications} with 
    preserved interactions.
A mixed-grained specification consists of {\it fine-grained} specifications
    of target modules to model code-level behavior
    and {\it coarse-grained} specifications of other modules to save cost.
Mixed-grained specifications enable us to 
    verify the system module by module~\cite{abadi1993composing-spec, abadi1995conjoining-spec, jonsson1994compositional-spec, clarke1989compositional-model-checking},
    and to verify code changes or bug fixes.

We write our specifications 
    in \tla{} which offers
    inherent flexibility to choose and adjust the abstraction level.
We present the principles of writing 
    multi-grained specifications with composability (\S\ref{sec:refinement}--\S\ref{sec:compose}).
We use conformance checking (\S\ref{sec:conformance})
    to match specifications with code implementation.

Concretely, use cases of multi-grained specifications are:
\begin{itemize}[leftmargin=*]
    \item {\bf Verifying protocol designs.}
    We verify the Zab algorithm
        using the protocol specification (\S\ref{sec:background:proto-spec}).
    As the protocol specification models high-level algorithms,
        it is verified in a traditional way without mixed-grained specifications.
    \item {\bf Verifying system designs.}
    As discussed in \S\ref{sec:zk_sys_spec},
        the system specification could take a long time to check, especially
        with complex configurations. 
    Mixed-grained specifications help the model checker
        speed up the verification.
    \item {\bf Verifying system implementations.}
    Mixed-grained specifications enable fine-grained modeling
        of code behavior to verify the implementation.
    The cost of model checking is managed by coarsening the modules 
        that are not verification targets.
    Conformance checking is needed to
         ensure specifications 
         conform to code implementations.
    \item {\bf Verifying code changes.}
    Mixed-grained specification also allows efficient
        verification of code changes (e.g., bug fixes).
    As code changes are typically local and incremental,
        we can use fine-grained specifications for the changed modules
        while coarsening the unchanged ones.
\end{itemize}

\subsection{Fine-Grained Specifications}
\label{sec:refinement}

For a given specification, we write its fine-grained counterpart by 
    modeling low-level code behavior.
For a target action in the original specification,
    we rewrite the enabling conditions based on code logic 
    and next-state updates of the action.
We focus on three patterns of model-code gaps (\S\ref{sec:background:gaps}):

\begin{itemize}[leftmargin=*]
    \item \textbf{Atomicity.} 
    We rewrite an action 
        that is not guaranteed to be atomically executed at the code level.
    We split the action 
        into multiple separate actions in the specification
        and set up their enabling conditions accordingly.
    \item \textbf{Concurrency.} 
    To model concurrency, 
        we separate state transitions, which are executed by different threads,
        into different actions as per the executing threads. 
    Inter-thread communications (e.g., message queues for local thread messages) 
        are also modeled in the specification. 

    \item \textbf{Missing state transitions.} 
    We focus on enhancing state transitions with existing variables in the target action.
    If the enabling condition includes other dependent variables that are missed,
        we add them to the specification.
\end{itemize}

\revision{We decide atomic blocks based on how threads/nodes communicate, 
    following prior work~\cite{hawblitzel2015ironfleet}. An atomic block 
    starts with reading the external state, performing internal computations, 
    and ends with writing to the external state. 
    The results of internal computations are invisible to other threads/nodes, 
    thus can be safely folded in an atomic block. An example atomic block starts with receiving 
    a message from the network, and ends with putting another message in a queue to be handled by another thread.}

\begin{figure}[htbp]	
    \begin{subfigure}[htbp]{0.495\textwidth}
        \includegraphics[width=\textwidth]{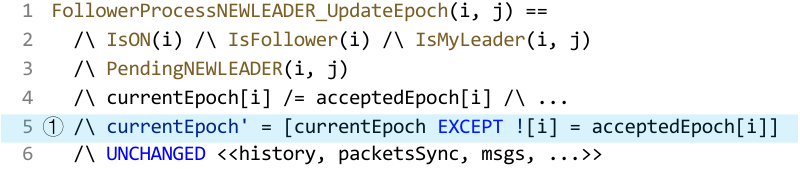}
        \caption{\bf Action 1: Updating the current epoch}
        \label{subfig:refine-part1}
    \end{subfigure}
    \begin{subfigure}[htbp]{0.495\textwidth}
        \includegraphics[width=\textwidth]{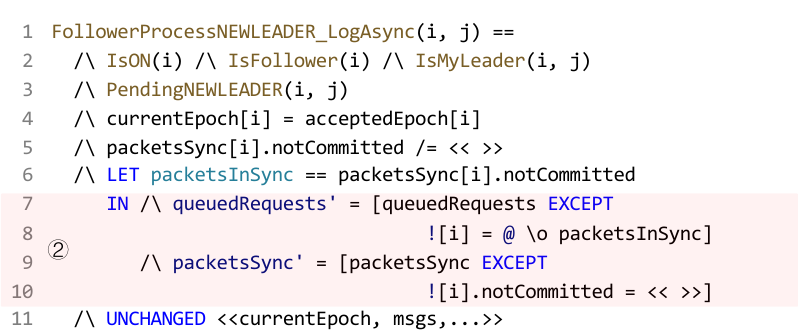}
        \caption{\bf Action 2: Queuing requests for asynchronous logging}
        \label{subfig:refine-part2}
    \end{subfigure}
    \begin{subfigure}[htbp]{0.495\textwidth}
        \includegraphics[width=\textwidth]{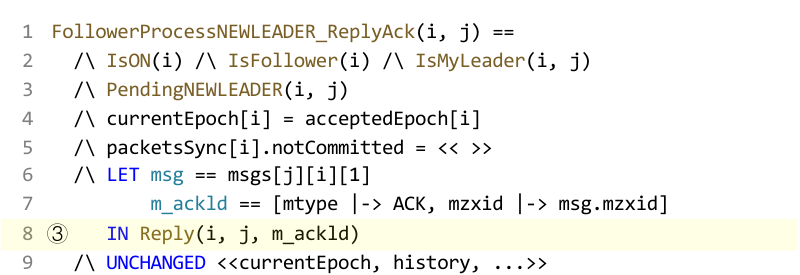}
        \caption{\bf Action 3: Sending \texttt{\small ACK} to the leader}
        \label{subfig:refine-part3}
    \end{subfigure}
    \caption{\bf Fine-grained modeling that splits the atomic action,
        \texttt{\small FollowerProcessNEWLEADER} in Figure~\ref{subfig:zab1.0-spec}, 
        into three actions (Actions 1--3).}
        \vspace{-10pt}
    \label{fig:refinement}
\end{figure}

\begin{figure}[htbp]
    \centering
    \begin{subfigure}[htbp]{0.495\textwidth}
        \includegraphics[width=\textwidth]{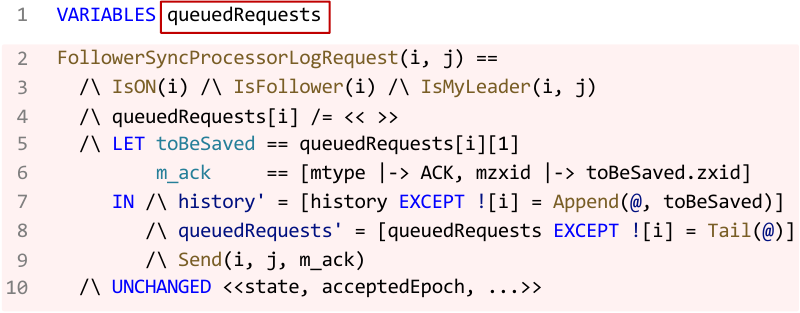}
        \vspace{-10pt}
        \caption{\bf Fine-grained specification in \tla{}}
        \vspace{-15pt}
        \label{subfig:syncProcessor-spec}
    \end{subfigure}
    \begin{subfigure}[htbp]{0.495\textwidth}
        \includegraphics[width=\textwidth]{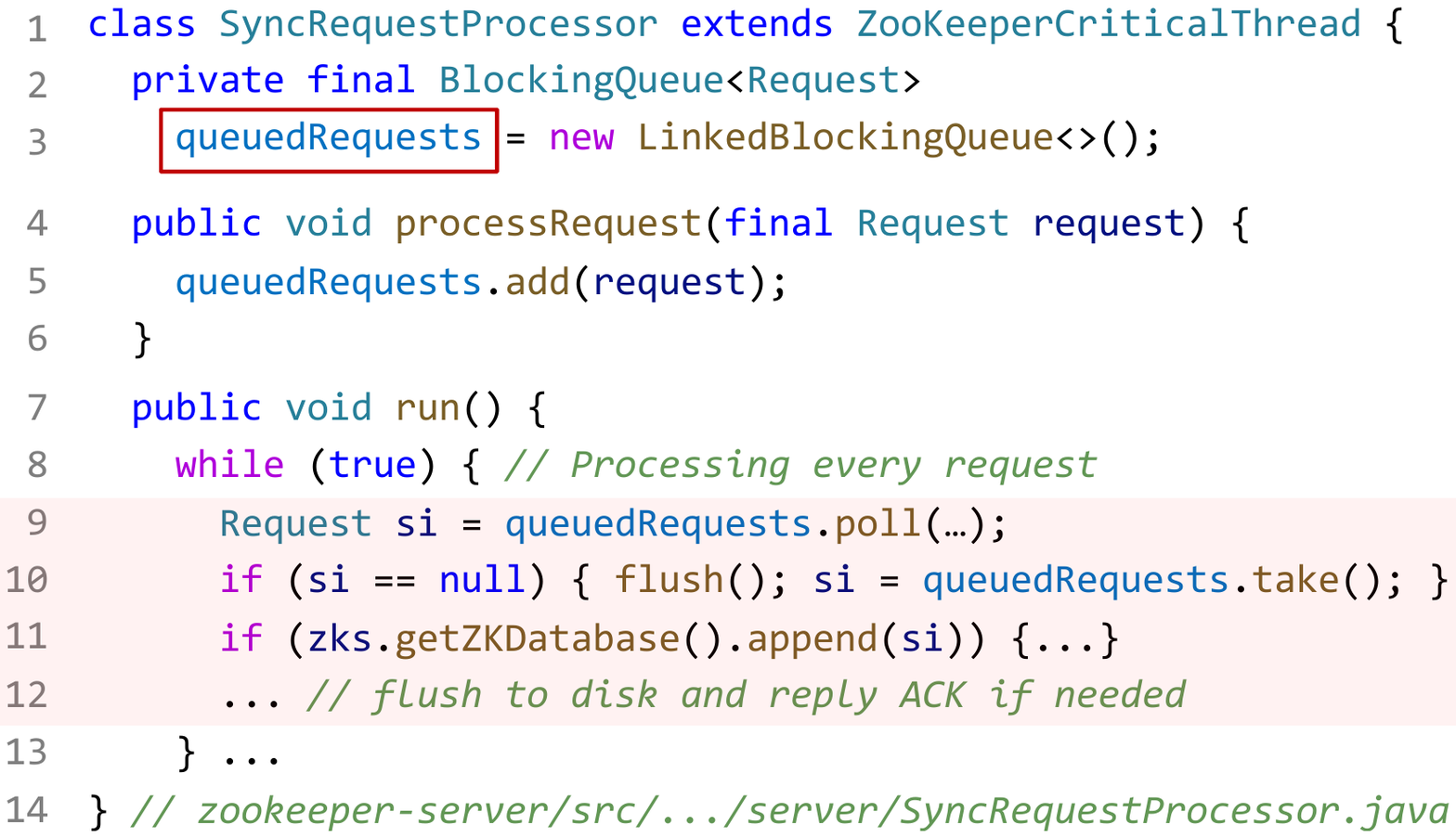}
        \vspace{-30pt}
        \caption{\bf Code implementation in Java}
        \label{subfig:syncProcessor-code}		
    \end{subfigure}
    \caption{{\bf Fine-grained modeling on concurrency with async logging.}
        \texttt{\small queuedRequests} is used in Figure~\ref{subfig:refine-part2}.}
    \vspace{-5pt}
    \label{F:syncProcessor-code-spec}
\end{figure}

\subsubsection*{Case study: Fine-grained modeling of Step $f.2.1$ in Figure~\ref{subfig:zab1.0-spec}} 
\label{subsec:refine-spec}
We rewrite the system specification 
    to model non-atomic actions and local concurrency
    with fine-grained specifications.
We first rewrite the atomic \texttt{\small FollowerProcessNEWLEADER}
    action into three actions corresponding to 
    steps \ding{172}, \ding{173}, and \ding{174} in Figure~\ref{subfig:zab1.0-spec},
    as there is no atomicity guarantee at the code level.
The fine-grained model allows model checkers to explore 
    intermediate states in non-atomic executions.
Figure~\ref{fig:refinement} shows the three actions in 
    the refined \tla{} specification.
    
To set up correct triggering of these fine-grained actions,
    we specify their enabling conditions using existing or new variables
    based on the code implementation.
For example, action \texttt{\small FollowerProcessNEWLEADER\_LogAsync} 
    models the logic of queuing requests 
    for asynchronous logging (Figure~\ref{subfig:refine-part2}).
It is enabled when the follower has updated its \texttt{\small currentEpoch} (line 4) 
    and there exist packets to be logged (line 5), 
    which corresponds to the conditions in the code
    (Figure~\ref{subfig:zab1.0-code}).

We then rewrite the \texttt{\small FollowerProcessNEWLEADER\_LogAsync}
    action for concurrency by specifying the asynchronous logging logic.
To do so,
    we decouple the logging action by a separate thread
    from the follower's message-handling actions, and make them interact properly. 
First, we change the variable for passing requests from the message-handling actions 
    (line 12-13 in Figure~\ref{subfig:zab1.0-spec})
    to a logging action (line 7-8 in Figure~\ref{subfig:refine-part2}).
We then model the logging action of the thread 
    and refactor the message-handling actions 
    to make them interact with the logging action.

Figure~\ref{subfig:syncProcessor-spec} shows the modeling of the asynchronous logging,
    including an additional variable \texttt{\small queuedRequests} 
    and a new action \texttt{\small FollowerSyncProcessorLogRequest}.
The variable \texttt{\small queuedRequests} 
    models the implemented queue (line 2-3 in Figure~\ref{subfig:syncProcessor-code})
    that stores the requests to be logged (line 4-6 in Figure~\ref{subfig:syncProcessor-code}).
The action \texttt{\small FollowerSyncProcessorLogRequest} (line 2-10 in Figure~\ref{subfig:syncProcessor-spec}) 
    focuses on the logic of logging requests (line 9-12 in Figure~\ref{subfig:syncProcessor-code}).
It takes out a request from \texttt{\small queuedRequests}, 
    logs it to disk and sends \texttt{\small ACK} to the leader.
\revision{In this way, we are able to model the actions that are concurrently executed by different threads in a node,
    for example,
    \texttt{\small FollowerSyncProcessorLogRequest} (Figure~\ref{subfig:syncProcessor-spec}).
    The interleavings of these locally concurrent actions can then be explored at the model level.
}

\subsection{Coarse-Grained Specifications}
\label{sec:abstraction}
    
When model checking a target module \revision{with the fine-grained specification}, 
    the other modules are coarsened 
    to reduce state space 
    and avoid state-space explosion.
To ensure verification safety,
    the coarsening must follow the {\it interaction preserving} principle,
    i.e., for each module, only the internal part can be omitted 
    while the interactions with other modules must be preserved,
    such that another module cannot distinguish whether it is interacting 
    with the original or a coarsened module.
This indistinguishability ensures the correctness of compositional model checking.

Safe coarsening is done by \revision{following} 
    the rationale of interaction preservation in \tla{}.
In \tla{}, we define the global states of a distributed system 
    with \textit{variables} and update the states with \textit{actions}.
We define \textit{dependency variables} of an action as 
the variables in the enabling condition of the action;
    the dependency relation is transitive---if a dependency 
    variable is calculated from another variable, 
    that variable is also a dependency variable.
The dependency variables of a module hence consist of dependency variables
    of all the actions in the module, where
    \revision{a \textit{module} is a set of actions.}
We define {\it interaction variables} as dependency variables shared by two modules.\footnote{Our definition 
        of \revision{interaction} variables is conservative, because
        dependency variables in two modules may not convey 
        any interaction.
    In practice, this case is rare 
        so we make the definition concise and easy to use.}

The coarsening preserves interaction if: (1) all dependency variables of the target module,
as well as all interaction variables, remain unchanged after the coarsening;
(2) all the updates of the dependency variables and interaction variables remain unchanged after the coarsening.
\revision{
These two constraints relate actions of a fine-grained module to those of the coarsened module 
    through dependency variables and interaction variables along with their updates.
}

We denote a specification $S$ that consists of $n$ modules as $S = \bigcup_{1 \le i \le n}M_{i}$,
and $\widetilde{M_i}$ as a module obtained by coarsening $M_i$ following the above two constraints.
\revision{
$S_i$ is denoted as the specification by coarsening every other module except $M_i$, i.e.,
 $S_i = (\bigcup_{j \neq i}\widetilde{M_j}) \cup M_i$.}

Let the traces allowed by $S$ and $S_i$ be $T_S$ and $T_{S_i}$ respectively.
When we are only concerned with the states of the target module $M_i$, 
    all traces in $T_S$ and $T_{S_i}$ are projected to $M_i$, 
    which are denoted as $T_S|_{M_i}$ and $T_{S_i}|_{M_i}$.
Then we can talk about the equivalence relation between traces with respect to a target module,
    which is defined as: $T_S \stackrel{M_i}{\sim} T_{S_i} \xlongequal{def} T_S|_{M_i} = T_{S_i}|_{M_i}$.

The safety of the coarsening is captured by the equivalence between traces, as in the following theorem.

\begin{namedtheorem}[Interaction Preservation]
Given $S = \bigcup_{1 \le i \le n}M_{i}$ and $S_i = (\bigcup_{j \neq i}\widetilde{M_j}) \cup M_i$,
we have $T_S \stackrel{M_i}{\sim} T_{S_i}$.
\end{namedtheorem}

\revision{Appendix~\ref{sec:appendix} provides the proof sketch of the theorem.}

\revision{
The key concept of the theorem is inspired by \cite{gu2022compositional} but is used differently.
In \cite{gu2022compositional}, interaction preservation is used 
    to establish the abstraction-refinement relation between different levels of specifications.
In this work, coarse-grained and fine-grained specifications 
    do not have abstraction-refinement relations due to model-code gaps.
Therefore, coarsening does not enforce abstraction relations. 
Both fine-grained and coarse-grained specifications are checked against implementation for conformance.
The correctness of modules under verification is guaranteed by ensuring invariants of system design.
(The correctness is not guaranteed by abstraction relations between fine-grained and the coarse-grained specifications as in \cite{gu2022compositional}.)
The interaction preservation ensures that all possible behaviors of the target module under verification is systematically explored, 
without noticing that internal details of its interacting modules are omitted.}

In our experience, identifying interaction variables in \tla{} specifications is 
    straightforward, 
    especially for systems designed with modularity and loose coupling.
We can also potentially borrow ideas from implementation-level model checking~\cite{guo2011practical}
    to dynamically identify
    interaction.

\subsubsection*{Case study: Coarsening the model of the Election and Discovery phases} 
\label{subsec:abstract-spec}
\begin{figure}	
    \centering
    \begin{subfigure}[htbp]{0.495\textwidth}
        \includegraphics[width=\textwidth]{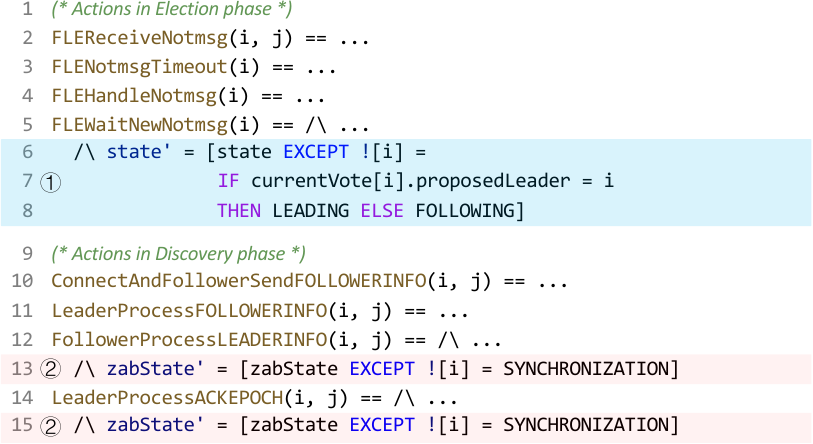}
        \caption{\bf Before coarsening (eight actions in two modules)}
        \label{subfig:abstract-before}		
    \end{subfigure}
    \begin{subfigure}[htbp]{0.495\textwidth}
        \includegraphics[width=\textwidth]{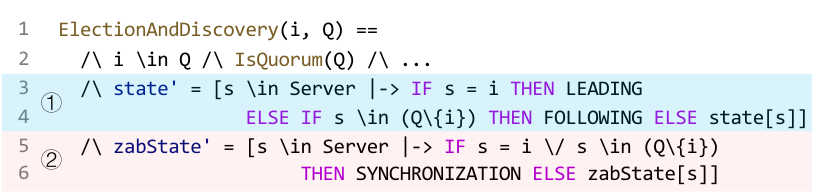}
        \caption{\bf After coarsening (one action)}
        \vspace{-10pt}
        \label{subfig:abstract-after}
    \end{subfigure}
    \caption{\bf Interaction-preserving coarsening of the eight actions 
        in the Election and Discovery phases.}
    \label{F:abstraction}
    \vspace{-10pt}
\end{figure}

In ZooKeeper's system specification, 
    the Election and Discovery phases are modeled by eight 
    atomic actions (Figure~\ref{subfig:abstract-before}).
An action may handle an incoming message (e.g., \texttt{\small LeaderProcessACKEPOCH}), 
    send a message to a peer (e.g., \texttt{\small ConnectAndFollowerSendFOLLOWERINFO}), 
    or broadcasts messages to peers (e.g., \texttt{\small FLEHandleNotmsg}).
If these two phases are not the target of the verification,
    we can coarsen the eight actions
    into one action, 
    as shown in Figure~\ref{subfig:abstract-after}.
    
To do so, we first identify internal variables that do not interact and thus do not affect other phases.
For example, one internal variable is \texttt{\small currentVote} (line 7 in Figure~\ref{subfig:abstract-before}).
This variable stores the leader information which is only consumed by the local node;
    hence,
    this variable can be abstracted away in the coarse-grained action.

In comparison, 
    variables \texttt{\small state} and  \texttt{\small zabState} 
    have external effects; hence, they cannot be abstracted away.
We use \ding{172} to show how the variable \texttt{\small state} 
    is updated before and after the coarsening.
Before coarsening, the \texttt{\small state} of a node is updated according to the leader information, 
    which is stored in \texttt{\small currentVote}.
After coarsening, all the participating nodes atomically update their \texttt{\small state} 
    as either \texttt{\small LEADING} or \texttt{\small FOLLOWING}  (line 3-4 in Figure~\ref{subfig:abstract-after}). 
This coarsening is interaction preserving, which is checked based on the constraints.
We use \ding{173} to \revision{show} how the variable \texttt{\small zabState} (i.e., the node phase) 
    is updated before and after the coarsening.  
Before the coarsening, \texttt{\small zabState} is updated in different actions (lines 13 and 
    15 in Figure~\ref{subfig:abstract-before});
after coarsening, all the participating nodes are collectively transitioned 
    to the Synchronization phase (line 5-6 in Figure~\ref{subfig:abstract-after}).

\subsection{Composition}
\label{sec:compose}

Composing modules and their actions is naturally supported by \tla{}.
The common practice is to define a next-state action that
nondeterministically chooses one action from one module
to run in each step.

\begin{figure}
    \centering
    \includegraphics[width=0.9\linewidth]{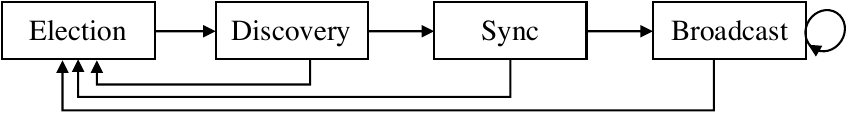}
    \caption{\textbf{We define four modules and write specifications
        per module based on phases in Zab.}}
    \vspace{-5pt}
    \label{fig:module}
\end{figure}

\begin{figure}
    \centering
    \includegraphics[width=\linewidth]{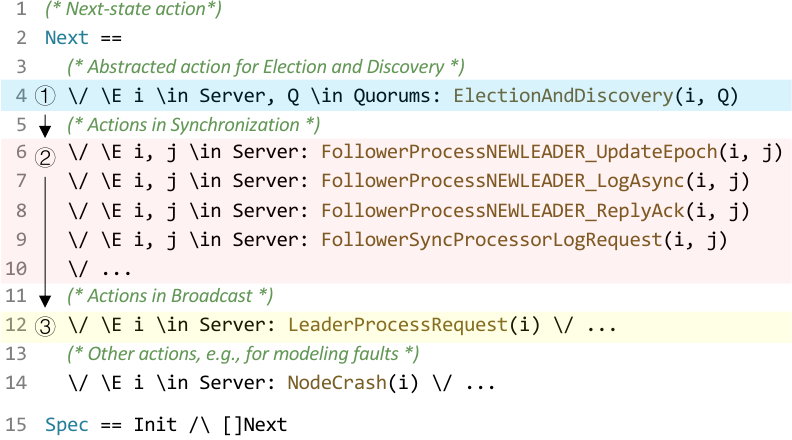}
    \vspace{-15pt}
    \caption{\textbf{Composition of coarse- and fine-grained actions.}
    \ding{172}: the coarsened action for Election and Discovery
    (Figure~\ref{subfig:abstract-after}).
    \ding{173}: the fine-grained actions for Synchronization
    (Figure~\ref{fig:refinement} and Figure~\ref{subfig:syncProcessor-spec}).
    \ding{174}: the actions for Broadcast.}
    \vspace{-5pt}
    \label{F:composition}
\end{figure}

We define four modules corresponding to the four phases of the Zab protocol 
    and compose their specifications into mixed-grained specifications.
The four phases are Election, Discovery,
Synchronization and Broadcast as shown in Figure~\ref{fig:module}.
Figure~\ref{F:composition} shows one example composition, including one (coarsened)
action for both Election and Discovery, (fine-grained) actions for Synchronization,
    and actions for Broadcast.
In addition, the next-state action also includes actions for modeling faults (e.g., a node crash).
The entire specification is defined as the initial state (\texttt{\small Init})
and the state transition represented by the next-state action (\texttt{\small []Next}).
Note that \texttt{\small []} is the temporal operator $\Box$ which, in this context, means
that the next-state action keeps running forever.

This style of composition captures non-deterministic natures of distributed systems.
In each step, \texttt{\small Next} chooses any possible action to run
since it is defined as the disjunction (\texttt{\small \textbackslash/})
of all actions.
If the action involves certain leader or follower node,
\texttt{\small Next} also chooses any possible node
using the existential quantifier (\texttt{\small {\textbackslash}E}).
So, the definition of \texttt{\small Next} allows
any (enabled) action with any server to happen at any point.

In this way, we compose different combinations of coarse- and fine-grained
modules into different specifications for checking different phases of the Zab implementation.

\subsection{Conformance Checking} 
\label{sec:conformance}

We ensure that all the \tla{} specifications we wrote, despite their granularities, 
    match the implementation through conformance checking.
For a specification, 
    the conformance checker explores model-level state space
    to generate traces, replays them in the implementation,
    and compares model- 
    and code-level execution traces (\S\ref{sec:impl:conformance}).

If the conformance checking detects a discrepancy between the model
    and the implementation,
    we debug the discrepancy and revise the specification to match code behavior.
The deterministic replay provided by our conformance checker makes it easy to 
    debug at the code level.
After the specification is updated, we run a new round of conformance checking
    until it passes.
The continuous conformance checking helped us find several discrepancies,
    ranging from inconsistent message types between 
    specification and implementation,
    to incorrect conditional branches and wrong variable assignments that 
    lead to unrealistic state transitions.

\revision{
\subsubsection*{Limitation.} 
The conformance checking is unsound.
It could miss behaviors in implementation that are not modeled by the specification, 
    while the discrepancy is not detected during conformance checking.
We consider improving it with 
    refinement checking and abstraction mapping~\cite{tasiran2003using,elmas2005vyrd}.
We only check safety properties instead of liveness properties 
    (which are hard for conformance checking).}

\subsection{\tool{}: Tooling Support} 
\label{sec:tool}

We wrote \tool{} as a framework and tooling support
    for creating mixed-grained specifications by composing 
    multi-grained specifications of different modules,
    interfacing the model checker (TLC),
    and providing conformance checking.

\subsubsection{Workflow}
We select the specification of each module
    and compose them into a mixed-grained specification
    for the entire system.
We write the specifications in a composable manner (\S\ref{sec:compose})
    so composing them is straightforward.
Currently, the selection is done manually;
future work can automate it based on time budget
    of model checking.
If there is no specification at the desired granularity,
    one can write a new specification (see \S\ref{sec:refinement} and \S\ref{sec:abstraction}).
The new specification will then be added into \tool{}.
In case that the specifications are not composable,
    parsing or semantic errors will be reported by the TLC model checker.

With multi-grained specifications, a specification is associated
    with protocol- and code-level invariants (\S\ref{subsec:inv}).
\tool{} automatically selects invariants when
    composing specifications,
    which will be checked during model checking.

We run continuous conformance checking to ensure the specification
    in \tool{} is synchronized with the implementation (\S\ref{sec:conformance}).
We describe our conformance checker which is built on top of 
    deterministic execution.

\subsubsection{Conformance checker}
\label{sec:impl:conformance}
The conformance checker randomly explores the model-level state space
to obtain a set of traces under a predefined time budget (e.g., 30 minutes).
For each trace, the conformance checker deterministically replays it at the
code level (\S\ref{sec:impl:dexec}) and
reports a discrepancy if (1) a model-level variable
and its code-level counterpart have different values, or
(2) a model-level action's code-level counterpart, once enabled,
never takes place (in 50 seconds).

Our conformance checker is not guaranteed to detect all discrepancies through random exploration,
similar to prior work~\cite{bornholt2021lightweight, tang2024sandtable, cirstea2024validating}.
To avoid false alarms caused by discrepancies,
\tool{} deterministically replays each model-level trace that violates some safety property
to confirm the safety violation also happens in the implementation.

Our conformance checker also reports implementation bugs with obvious symptoms like assertion failures
when replaying traces. Developers can mark such traces and later during model checking
\tool{} will focus on exploring other traces since the marked traces are already known as buggy.

\subsubsection{Deterministic execution}
\label{sec:impl:dexec}

The conformance checker needs to deterministically replay model-level
traces at the code level.
\tool{} realizes deterministic execution by having a central coordinator
that intercepts and coordinates actions from different threads on different nodes.
The coordinator takes a model-level trace as the input, 
schedules the code-level actions one by one accordingly,
and injects faults (e.g., node crashes) when needed.

To deterministically replay a model-level trace, the coordinator needs to
map each model-level action to the code,
and precisely control the interleaving between code-level actions.
\tool{} currently requires developers to provide a mapping from each model-level action
to the events that represent the beginning and the end of the corresponding code-level action.
\tool{} then instruments around such events to control the interleaving.
For example, the model-level action \texttt{\small FollowerProcessNEWLEADER(f,l)}'s
corresponding code-level action begins with the leader \texttt{\small l}
calling the \texttt{\small writeRecord} method
to send a \texttt{\small \revision{NEWLEADER}} message to the follower \texttt{\small f},
and ends with \texttt{\small f} sending back an \texttt{\small ACK} message
(the \texttt{\small writeRecord} method is used for sending messages between ZooKeeper nodes).
Developers provide this mapping and then \tool{} instruments \texttt{\small writeRecord}
to inject an RPC client that calls the coordinator during runtime with the context information
(e.g., arguments and the caller of \texttt{\small writeRecord}).
The call returns only when the coordinator schedules this action according to the model-level trace.
The coordinator will not schedule any other actions
until the currently running action ends.
In this way, the coordinator deterministically decides when each code-level action begins
and controls the interleaving between code-level actions.

\revision{
If the implementation can interleave events 
    that appear atomic at the model level and generate discrepant states,
    developers must either revise the specification to enable the code-level interleaving,
    or provide a more precise mapping to make the conformance.
For example, the election messages can interleave during Election, 
    with a non-deterministic leader generated. 
For a coarsened Election action that elects a target leader at the model level, 
    developers can set the messages that vote for the target leader with higher priority. 
    In this way, the deterministic replay is able to generate a matched state required by the model-level action.
The debugging process is iterative until the specification and the mapping
    reach satisfactory conformance.
}

The deterministic execution is also useful for debugging safety violations.
For a trace that violates a safety property during model checking,
\tool{} deterministically reproduces it at the code level,
so that developers can diagnose the root cause of the safety violation.

\tool{} implements the deterministic execution coordinator using the Java Remote Method Invocation~\cite{rmi} framework
and instruments ZooKeeper using AspectJ~\cite{aspectj} to inject the RPC clients.
For each version of the specification, developers need to provide a mapping from each
model-level action to the corresponding code-level action.

An alternative option for realistic execution is to intercept system calls~\cite{tang2024sandtable}.
Intercepting system calls avoids system-specific instrumentation and leaves the target system unmodified.
However, we choose to instrument the target system because it is hard to control user-level threads
(e.g., event handlers in ZooKeeper) accurately at the system call level.

\section{Verifying ZooKeeper} 
\label{sec:lfm}

We wrote the protocol specification of Zab (\S\ref{sec:background:proto-spec}),
    as well as the system specification of ZooKeeper (\S\ref{sec:zk_sys_spec}),
    including the 
    specification of the Fast Leader Election (FLE)
    which was not a part of the original Zab protocol~\cite{junqueira2011zab}.
We use the TLC model checker to verify both of them.
We build on the system specification as the baseline to
    develop code specifications
    and compose mixed-grained specifications as discussed in \S\ref{sec:spec}.
The mixed-grained specifications are then used to verify the 
    implementation of ZooKeeper and its code changes.

\subsection{Mixed-Grained Specifications for Log Replication} 
\label{subsec:dev}

We present the mixed-grained specifications composed using \tool{}
    to verify the log replication of ZooKeeper.
We focus on log replication because it is 
    the main procedure for achieving consensus in ZooKeeper.\footnote{\revision{We also verify the other part of ZooKeeper and 
    find bugs including ZK-2776, ZK-3336, ZK-3707, ZK-4040, ZK-4416 and ZK-4781. These bugs are known bugs but still exist in the checked versions.}}
It involves both the Synchronization and Broadcast modules.
The log replication implementation 
    has been greatly optimized during the evolution of \zk;
hence, the code, especially the Synchronization module, can be described by neither the protocol nor the system specifications effectively.
As a result, severe bugs (e.g., those that lead to data loss or inconsistencies) 
    were constantly reported, such as~\cite{ZK-2355,ZK-2845,ZK-3023,ZK-3104,ZK-3911}.

Our first step is to ensure the system specifications
    match the ZooKeeper implementation.
We run conformance checking  and find a discrepancy where 
    the model-level traces cannot be successfully replayed 
    at the code level due to an unexpected exception.
The root cause is that at the code level,
    if the follower receives the \texttt{\small COMMIT} message {\it after} the \texttt{\small NEWLEADER} message, 
    it will throw an unexpected \texttt{\small NullPointerException} and terminate the Synchronization phase. 
The issue had been reported in ZK-4394~\cite{ZK-4394} but was not resolved.
So, we 
    adjusted the specification by adding new conditions 
    and a new commit assertion to check the \texttt{\small NullPointerException}
    and avoiding further exploration once ZK-4394 occurs.

Based on the conformed system specification, we create 
    the following mixed-grained specifications (Table~\ref{table:compo-spec}).

\begin{table}
    \footnotesize
    \setlength{\tabcolsep}{4pt}
    \begin{tabular}{lcccc}
        \toprule
        \multicolumn{1}{c}{} &  &  & \multicolumn{2}{c}{\textbf{Log   Replication}} \\ \cmidrule(lr){4-5} 
        \multicolumn{1}{c}{\multirow{-2}{*}{\textbf{Spec}}} & \multirow{-2}{*}{\textbf{Election}} & \multirow{-2}{*}{\textbf{Discovery}} & \textbf{Synchronization} & \textbf{Broadcast} \\
        \midrule
        SysSpec & \cellcolor{LightGray2}Baseline & \cellcolor{LightGray2}Baseline & \cellcolor{LightGray2}Baseline & \cellcolor{LightGray2}Baseline \\
        \mixedSpec{}-1 & \multicolumn{2}{c}{\cellcolor{VeryLightGray2}Coarsened} & \cellcolor{LightGray2}Baseline & \cellcolor{LightGray2}Baseline \\
        \mixedSpec{}-2 & \multicolumn{2}{c}{\cellcolor{VeryLightGray2}Coarsened} & \cellcolor{Gray2}{\color{black}Fine-grained (atom.)} & \cellcolor{LightGray2}Baseline \\
        \mixedSpec{}-3 & \multicolumn{2}{c}{\cellcolor{VeryLightGray2}Coarsened} & \cellcolor{Gray2}{\color{black}\begin{tabular}[c]{@{}c@{}}Fine-grained\\ (atom.+ concur.)\end{tabular}} & \cellcolor{Gray2}{\color{black}\begin{tabular}[c]{@{}c@{}}Fine-grained\\ (concur.)\end{tabular}} \\
        \mixedSpec{}-4 & \cellcolor{LightGray2}Baseline & \cellcolor{LightGray2}Baseline & \cellcolor{Gray2}{\color{black}\begin{tabular}[c]{@{}c@{}}Fine-grained\\ (atom.+ concur.)\end{tabular}} & \cellcolor{Gray2}{\color{black}\begin{tabular}[c]{@{}c@{}}Fine-grained\\ (concur.)\end{tabular}} \\
        \bottomrule
    \end{tabular}
    \caption{{\bf Mixed-grained specifications for verifying log replication, composed 
        from multi-grained specifications.}
        ``SysSpec'' refers to the system specification that passes
        conformance checking (used as the baseline).}
        \vspace{-10pt}
    \label{table:compo-spec}
\end{table}

\begin{table*}[t]
    \footnotesize
    \begin{tabular}{cll}
        \toprule
        \textbf{ID} & \textbf{Invariant(s)} & \textbf{Source} \\                                               
        \midrule
        I-1  & {\bf Primary uniqueness.} There is at most one established leader for each epoch. & Protocol \\
        I-2  & {\bf Integrity.} If some process delivers $t$, then some primary has broadcast $t$. & Protocol    \\
        I-3  & {\bf Agreement.} If some process $f$ delivers $t$, and some process $f'$ delivers $t'$, then $f'$ delivers $t$ or $f$ delivers $t'$.  & Protocol\\
        I-4  & {\bf Total order.} If some process delivers $t$ before $t'$, then any process that delivers $t'$ must also deliver $t$ and deliver $t$ before $t'$. & Protocol\\
        I-5  & {\bf Local primary order.} If a primary broadcasts $t$ before it broadcasts $t'$, then a process $f$ that delivers $t'$ must also deliver $t$ before $t'$. & Protocol \\
        I-6  & {\bf Global primary order.} If a process $f$ delivers both $t$ (in epoch $e$) and $t'$ (in epoch $e'$, $e < e'$), then $f$ must deliver $t$ before $t'$. & Protocol\\
        I-7  & {\bf Primary integrity.} If a primary $\rho_e$ broadcasts $t$ and some process $f$ delivers $t'$ st. $t'$ has been broadcast by $\rho_{e'}$, $e'<e$, & Protocol\\
             & \ \ \ \ then $\rho_e$ must deliver $t'$ before it broadcasts $t$. & \\
        I-8  & {\bf Initial history integrity.} Let $e$, $e'$ be epochs, $e < e'$, and $e$ be an established epoch. $I_e \sqsubseteq I_{e'}$. & Protocol\\
        I-9  & {\bf Commit consistency.} Let $\Delta_f$ be the delivered transaction sequence of process $f$, and $f.e$ be $f$'s last committed epoch. $I_{f.e} \sqsubseteq \Delta_f$. & Protocol\\
        I-10 & {\bf History consistency.} For any two processes $f$ and $f'$ that participate in epoch $e$, either $h_f \sqsubseteq  h_{f'}$ or $h_{f'} \sqsubseteq h_f$. & Protocol\\
        \midrule
        I-11  & {\bf Bad states (4 instances).} Exceptions or false assertions on the server states upon receiving certain types of messages. & Code \\
        I-12  & {\bf Bad acknowledgments (2 instances).} Exceptions or false assertions on the \texttt{\scriptsize ACK} message content processed by the leader. & Code \\
        I-13  & {\bf Bad proposals (2 instances).} Exceptions or false assertions on the \texttt{\scriptsize PROPOSAL} message content processed by the follower. & Code \\
        I-14  & {\bf Bad commits (3 instances).} Exceptions or false assertions upon handling the \texttt{\scriptsize COMMIT} message or committing a transaction. & Code\\
        \bottomrule
        \end{tabular}
    \caption{{\bf Invariants including safety properties of the Zab protocol and the code-level assertions by developers}. 
        $t$: a transaction; 
        $h_f$: (transaction) history of process $f$; 
        $\rho_{e}$: primary of epoch $e$; 
        $\sqsubseteq$: the relation of prefix;
        $I_e$: initial history of epoch $e$.}
    \vspace{-10pt}
    \label{table:invariants}
\end{table*}

\begin{itemize}[leftmargin=*]

\item {\bf \mixedSpec{}-1.}
Since we target log replication,
    we coarsen the Election and Discovery modules.
Section~\ref{sec:abstraction} described the coarsening which 
    coarsens the eight actions of the Election and Discovery modules 
    in the system specification 
    into one \texttt{\small ElectionAndDiscovery} action.

\item {\bf \mixedSpec{}-2.}
\revision{
    We write a fine-grained specification of}
    the Synchronization module to model the 
    non-atomic updates of epoch and history when a follower receives the \texttt{\small NEWLEADER} message (\S\ref{sec:refinement}).
So, the model checker can explore intermediate states between 
    the updates of epoch and history,
    which is induced by node crashes.
As we focus on log replication, 
    \mixedSpec{}-2 uses the coarsened action for the Election and Discovery modules
     in \mixedSpec{}-1.

\item {\bf \mixedSpec{}-3.}
This specification further models multithreading concurrency in log replication.
There are three main threads in the follower process,
    for handling incoming messages, 
    logging transactions, 
    and committing transactions, respectively.
We distinguish the asynchronous actions executed by different threads,
    and specify them into separate actions, as demonstrated in \S\ref{sec:refinement}.
\mixedSpec{}-3 also uses the coarsened \texttt{\small ElectionAndDiscovery}.

\item {\bf \mixedSpec{}-4.} As a reference, we create 
    \mixedSpec{}-4 by composing system specifications
    of the Election and Discovery modules 
    and fine-grained log replication modules in \mixedSpec{}-3.
\end{itemize}

\subsection{Invariants}  
\label{subsec:inv}

We specify 14 invariants that are checked throughout the model checking,
    as shown in Table~\ref{table:invariants}. 
Ten invariants are safety properties defined by the Zab protocol~\cite{junqueira2010dissecting},
    including both core properties and lemmata;
these invariants must be satisfied by any Zab implementations.
During model checking, 
    these ten invariants are checked upon state transition.
These invariants apply to specifications of any granularity.

We also define other four types of invariants (11 instances in total)
    based on the code-level implementation of ZooKeeper.
We observe that developers add additional checks on 
    specific behavior (e.g., by throwing exceptions and using assertions).
Some of them are not reflected by the safety properties of the 
    Zab protocol, and thus shall be included in fine-grained specifications.
Some of them are caused by known, but still not resolved 
    bugs in the code, such as ZK-4394 discussed 
    in \S\ref{subsec:dev}.
Essentially, each of these invariants specify 
    the execution is on an error path.
These invariants are checked whenever the model checker 
    reaches the corresponding execution path.
Note that code-level invariants are specific to
    certain granularities that model the corresponding execution.

\begin{table}
    \footnotesize
    \setlength{\tabcolsep}{3pt}
    \begin{tabular}{lccccc}
        \toprule
        \textbf{Specification diff.} & \textbf{Lines} &  \textbf{Variables} & \textbf{Actions}  & \textbf{Instr.}  & \textbf{Hour}\\
        \midrule
        \mixedSpec{}-1 $-$ SysSpec         & +64, -342   & 29 (-8) & 16 (-7)    & 31 (+0)  & 18  \\
        \mixedSpec{}-2 $-$ \mixedSpec{}-1  & +34, -19    & 29 (+0) & 17 (+1)    & 32 (+1)  & 8  \\
        \mixedSpec{}-3 $-$ \mixedSpec{}-2  & +188, -118  & 31 (+2) & 19 (+2)    & 36 (+4)  & 40 \\
        \bottomrule
    \end{tabular}
    \caption{\textbf{Efforts of writing multi-grained specifications.} 
        ``\#Instr.'' refers to the number of instrumentation pointcuts.
        }
    \vspace{-10pt}
    \label{table:cost}
\end{table}

\subsection{Efforts} 
\label{sec:effort}
Table~\ref{table:cost} shows the efforts of writing multi-grained specifications
    in Table~\ref{table:compo-spec}.
The efforts of writing and maintaining multi-grained 
    specifications are manageable, especially
    when baseline specifications are available.
Fine-grained modeling and coarsening can be done incrementally 
    on top of the reference specification.
For example, the differences between the specifications
    are less than 500 lines.
Following composable formal methods, all invariants and most variables in the baseline specification 
    are directly reused.
In addition, we need to provide a mapping from the newly added (coarse- or fine-grained)
    model-level actions to the code-level actions so that \tool{} can instrument code
    for deterministic execution (\S\ref{sec:impl:dexec}) at different granularities.
Overall, the effort is done within 40 person-hours,
    and is done by one person who is familiar with the ZooKeeper code 
    and is proficient in \tla{}.
Further, as more specifications are written, 
    the reusability of composable components grows higher,
    amortizing the cost.

\subsection{Setup}

We use a configuration of a three-node ZooKeeper cluster with up to four transactions,
    up to three node crashes, and up to three network partitions.
This configuration is a common practice used in prior work~\cite{leesatapornwongsa2014samc, lukman2019flymc,wang2023mocket,tang2024sandtable}.
We use TLC to run model checking on \tla{} specifications on a single machine.
We use TLC's breath-first search (BFS) as the strategy for state-space exploration.
With BFS, once an invariant is violated, 
    we can obtain the buggy trace with minimal depth.

\section{Results and Experience} 
\label{sec:eval}

\subsection{Verification Results} 
\label{subsec:effectiveness}

We start to systematically model check ZooKeeper using mixed-grained specifications
    since version v3.9.1.
ZooKeeper did not pass the verification.
The model checking exposes a total of six severe bugs, as shown in Table~\ref{table:bugs-found}.
All these bugs have serious consequences,
    including data loss, data inconsistencies, and data synchronization failures.
All these bugs are deep bugs, as their manifestations take minimal depths 
    of tens of actions and more than tens of thousands of states,
    which are hard for developers to reason about.
For the same reason, they are also hard to fix (\S\ref{sec:verify-fix}).

Those bugs are found when TLC reports violations of invariants
    in the traces during model checking.
We then confirmed the bugs by deterministically replaying the traces at the code level
    using \tool{} (\S\ref{sec:tool}).
To debug a violation, we analyze the model-level trace 
    to understand the triggers and locate 
    the root cause in the code.
The debugging is eased by traces with minimal depth explored by the BFS strategy.
In practice, we start the model checking from a small configuration (e.g., one node crash) 
    to a large one (up to three node crashes),
    which helps us obtain a simple and concise trace.

Table~\ref{table:bugs-found} shows the most efficient specification that found each bug.
All the bugs except one (ZK-4394) require \mixedSpec{}-2 and \mixedSpec{}-3.
In other words, finding these bugs needs fine-grained modeling of non-atomic actions
    and local concurrency; these bugs cannot be found 
    by the baseline system specification (\S\ref{sec:zk_sys_spec}).
The results show the importance of closing the model-code gap 
    with fine-grained specifications that reflect 
    code-level behavior.
We discussed ZK-4394 in \S\ref{subsec:dev}. 
It can be found by \mixedSpec{}-1
    with conformance checking.
The bug can be found with the system specification also;
but, compared with the system specification, 
    \mixedSpec{}-1 significantly reduces the 
    time to find the bug.
In fact, all the bugs were found in less than two minutes
    with the benefits of specification coarsening. 
We discuss more about the efficiency of coarse-grained specifications in \S\ref{subsec:efficiency}.

\begin{table}
    \footnotesize
    \setlength{\tabcolsep}{2pt}
    \begin{tabular}{lllcccc}
    \toprule
    \textbf{Bug ID} & \textbf{Impact} 
                            & \textbf{Spec.}  & \textbf{Time} & \textbf{Depth} & \textbf{\#States} & \textbf{Inv.} \\ 
    \midrule
    \hrefzkissue{3023} & Data sync failure  & \mixedSpec{}-3 & 11 sec  & 13  & 78,892  & I-11    \\ 
    \hrefzkissue{4394} & Data sync failure  & \mixedSpec{}-1* & 9 sec   & 20  & 14,264  & I-14    \\ 
    \hrefzkissue{4643} & Data loss           & \mixedSpec{}-2 & 17 sec  & 21  & 208,018 & I-8   \\ 
    \hrefzkissue{4646} & Data loss           & \mixedSpec{}-3 & 109 sec & 21  & 2,880,498 & I-8 \\ 
    \hrefzkissue{4685} & Data sync failure   & \mixedSpec{}-3 & 10 sec  & 12  & 67,418    & I-12    \\ 
    \hrefzkissue{4712} & Data inconsistency & \mixedSpec{}-3 & 11 sec  & 13  & 73,293   & I-10    \\ 
    \bottomrule
    \end{tabular}
    \caption{\textbf{Bug detection in ZooKeeper v3.9.1.} 
    ``Spec.'' shows the most efficient mixed-grained specification to find the bug.
    ``Inv.'' shows the first violated invariant triggered by the bug.
    ZK-4394 is not fixed yet so we masked it in our specifications;
    \mixedSpec{}-1* refers to the specification before masking it.} 
    \vspace{-15pt}
    \label{table:bugs-found}
\end{table}

Three of the six bugs violate protocol-level invariants 
    and the others violate code-level invariants.
We find that code-level invariants are important,
    as developers often directly throw exceptions to abort 
    the execution in certain critical cases, 
    which is not defined in the protocol (\S\ref{subsec:inv}).

\revision{Appendix~\ref{sec:bugs} describes these bugs in more details.}

\subsection{Efficiency} 
\label{subsec:efficiency}

Mixed-grained specifications effectively improve efficiency
    of verifying the target modules.
We evaluate the verification efficiency 
    of the five specifications in Table~\ref{table:compo-spec}.
We set the time budget to be 24 hours 
    and the violation limit to be 10,000.
We use each of these specifications to verify ZooKeeper v3.7.0.
We run TLC (v1.7.0) with the BFS mode for exploration.
All the experiments were run on an Ubuntu 22.04 server with
    two AMD EPYC 7642 processors at 3.3GHz;
    each processor has 48 cores and 96 hyperthreads.
Each specification is checked by 16 workers (threads) with 32 GB memory.

\begin{table}
    \footnotesize
    \begin{subtable}[t]{\linewidth}
        \centering
        \begin{tabular}{lcccc}
            \toprule
            {\textbf{Spec}} & \textbf{Time} & \textbf{Depth} & \textbf{\# States} & \textbf{\# Violated Inv.}  \\
            \midrule
            {Baseline}        & >24h     & 26   & 2,271,335,268 & None  \\ 
            {\mixedSpec{}-1}  & 12m20s   & 56   & 17,586,953    & None  \\
            {\mixedSpec{}-2}  & 1m15s    & 21   & 2,237,960     & I-8  \\
            {\mixedSpec{}-3}  & 11s      & 13   & 77,179        & I-10   \\
            {\mixedSpec{}-4}  & 8h32m6s & 24   & 967,810,552    & I-10   \\
            \bottomrule
        \end{tabular}
        \caption{\textbf{Stopping at the first violation}}
        \label{tab:efficiency:mode-a}
        \end{subtable}
    \begin{subtable}[t]{\linewidth}
        \centering
        \setlength{\tabcolsep}{2pt}
        \begin{tabular}{lccccc}
            \toprule
            {\textbf{Spec}}  & \textbf{Time} & \textbf{Depth} & \textbf{\# States} & \textbf{\# Violation} & \textbf{\# Vio. Inv.} \\
            \midrule
            {Baseline}           & >24h     & 26    & 2,271,335,268    & 0         & None    \\
            {\mixedSpec{}-1}     & 12m20s   & 56    & 17,586,953       & 0         & None     \\
            {\mixedSpec{}-2}     & 15m55s   & 62    & 24,211,064       & 1,404     & I-8       \\
            {\mixedSpec{}-3}     & 5m10s    & 21    & 1,727,234        & >10,000   & I-10, I-11, I-12 \\
            {\mixedSpec{}-4}     & >24h     & 26    & 2,478,453,900    & 35        & I-10, I-11, I-12 \\
            \bottomrule
        \end{tabular}
        \caption{\textbf{Running to completion}}
        \label{tab:efficiency:mode-b}
    \end{subtable}
    \caption{\textbf{Verification efficiency of specifications with different granularities.}
    The configuration is three servers, two transactions, two crashes, and two partitions.
    ``Depth'' refers to the number of state transitions;
    ``States'' refers to the {\it distinct} states explored (and reported) by TLC.}
    \vspace{-10pt}
    \label{table:efficiency}
\end{table}

Table~\ref{table:efficiency} shows verification efficiency results in two modes:
    (\ref{tab:efficiency:mode-a}) stopping at the first violation,
    and (\ref{tab:efficiency:mode-b}) running to completion (till the limit).    
The mixed-grained specifications with fine-grained modeling
     (\mixedSpec{}-2, -3, and -4) 
    detect violations within the time limit.
The baseline (system specification) and \mixedSpec{}-1 find no violation 
    because of not modeling fine-grained behavior.
The baseline and \mixedSpec{}-4 cannot finish in 24 hours,
and we observe that TLC spends most of the time in the Election
    module without reaching other modules
    (the leader election algorithm is complex and takes many steps).
\mixedSpec{}-4 costs 2793$\times$ more time to detect the first violation 
    compared to \mixedSpec{}-3 as it does not coarsen the Election and Discovery modules.
In comparison, \mixedSpec{}-1, -2, and -3 
    coarsen the Election and Discovery modules,
    enabling TLC to more efficiently check the log replication modules.
    As a result, these three mixed-grained specifications 
    can finish in tens of minutes, 
    with \mixedSpec{}-2 and -3 
    finding the first violation in minutes.
So, mixed-grained specifications 
    provide the flexibility to help the model checker 
    focus on target modules with fine-grained models,
    which is critical to finding bugs and receiving prompt feedback.

\subsection{Verifying Bug Fixes}
\label{sec:verify-fix}

\begin{figure}
    \vspace{-7.5pt}
   \centering
    \def\svgwidth{0.85\linewidth}
    \small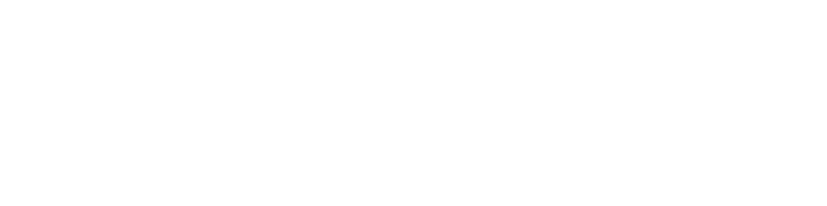\hspace*{-1.5em}
    \vspace{-5pt}
   \caption{{\bf Bugs introduced in ZooKeeper's log replication implementation.}
    * refers to those with fixes merged.} 
   \label{F:bug-history}
\end{figure}

Fixing bugs is challenging---it is easier to prevent specific symptoms, 
    but harder to rule out root causes due to the complexity 
    of reasoning about all interleavings of actions.
We first fixed ZK-4712, but found the other bugs are much harder to fix.
They are rooted in various performance optimizations since 2017, triggered by ZK-2678~\cite{ZK-2678}.
These optimizations have introduced over ten data loss/inconsistency bugs~\cite{ZK-2845,ZK-3023,ZK-3642,ZK-3911,ZK-4394,ZK-4541,ZK-4643,ZK-4646,ZK-4685,ZK-4712,ZK-4785}.
Some of them were fixed, while 
    others were not.
Without sufficient verification at the time,
    some accepted fixes introduced new bugs (Figure~\ref{F:bug-history}).

Mixed-grained specifications enable us to verify 
    bug fixes efficiently.
We verify four Pull Requests (PRs) 
    that attempted to fix the bugs in Table~\ref{table:bugs-found}.
All the PRs use multithreading with 
    non-atomic updates of epoch and history in the Synchronization phase.
Therefore, we use the specification of \mixedSpec{}-3 as the base specification.
With the fix of ZK-4712, we updated \mixedSpec{}-3, referred to 
    as \mixedSpec{}-3$^+$ (verified using TLC).
For each PR, we map the code changes to \mixedSpec{}-3$^+$
    and update \mixedSpec{}-3$^+$ accordingly.
Table~\ref{table:verify-fix} shows the verification results     
    with the mode of stopping at the first violation.
All four fixes are detected with invariant violations
    within five minutes.

\begin{table}[t]
    \footnotesize
    \setlength{\tabcolsep}{4pt}
    \begin{tabular}{lcccccc}
        \toprule
        ~   & \textbf{Spec.} & \textbf{Change} & \textbf{Time} & \textbf{Depth} & \textbf{\#States} & \textbf{Inv.}  \\
        \midrule
        {\hrefzkpr{1848}} & \mixedSpec{}-3$^+$ & +68, -29   & 274s & 21 & 8,166,775 & I-8   \\
        {\hrefzkpr{1930}} & \mixedSpec{}-3$^+$ & +102, -66  & 17s & 13 & 270,881    & I-12  \\
        {\hrefzkpr{1993}} & \mixedSpec{}-3$^+$ & +71, -51   & 34s & 15 & 765,437    & I-11  \\
        {\hrefzkpr{2111}} & \mixedSpec{}-3$^+$ & +70, -43   & 38s & 15 & 808,697    & I-11  \\ 
        \bottomrule
    \end{tabular}
    \caption{\textbf{Verifying bug fixes (pull requests).} 
            \mixedSpec{}-3$^+$ is the specification of \mixedSpec{}-3 with the fix of ZK-4712.}
    \vspace{-20pt}
    \label{table:verify-fix}
\end{table}

The challenge of fixing these bugs is that all of them
    involve the logic of handling the \texttt{\small NEWLEADER} message. 
So, it is hard to fix them in isolation---the fix of one bug 
    can make it hard to fix the others.
For example, PR-1993 targets ZK-4646 and ZK-4685, 
    without considering other bugs like ZK-4394.
To further fix ZK-4394 based on PR-1993 need
    heavy revision on the code logic of PR-1993.
If we consider the full picture,
    PR-1993 is not a good step towards a complete solution. 
Some fixes lead to new bugs, e.g., the merged fix 
    for ZK-3911 did not prevent the same violation,
    but opened new triggering paths of ZK-3023
    and induced new bugs like ZK-4685.

The mixed-grained specifications help us understand the root causes 
    and systematically verify whether a given code fix 
    is completed by exhaustively
    exercising all the possible interleavings modeled 
    by the fine-grained specification.
With a holistic understanding, 
    we developed a fix that resolves {\it all} the bugs in Figure~\ref{F:bug-history},
    and verify it with extensive model checking.
We discuss our resolution in \S\ref{subsec:improvement}.
The verified fixes have been merged to the latest version of ZooKeeper.

\subsection{Improving the Zab Protocol}
\label{subsec:improvement}

One essential reason for the error-proneness 
    of log replication
    is rooted in deviation
    of the implementation in ZooKeeper
    from the Zab protocol.
For example, the atomicity of updating epoch and history
    is explicitly required by the protocol,
    but is not followed by the implementations.\footnote{The Zab protocol implemented in ZooKeeper~\cite{Zab-wiki} 
    has evolved in several other ways different from 
    original papers~\cite{junqueira2010dissecting,junqueira2011zab}.}
Basically, the protocol no longer guides the implementation.
To fix existing bugs and also make it easy to implement correctly,
    we remove the atomicity requirement of the two updates
    from the Zab protocol but require 
    their order---the follower updates its history {\it before} updating its epoch.

We update the protocol specification (\S\ref{sec:background:proto-spec}),
    which splits the action of handling a \texttt{\small NEWLEADER} message 
    into two serialized actions of updating history
    and epoch.
In this way, the model checker can explore traces when a follower crashes right after it updates history. 
We add a variable \texttt{\small servingState} 
    to help express the enabling conditions of updating history and epoch.
We run extensive model checking with TLC to verify the new protocol specification
    and it passes all the ten protocol-level invariants (Table~\ref{table:invariants}).

We then update the implementation based on the new protocol,
    and address all the bugs in Figure~\ref{F:bug-history} 
    by enforcing the order of several critical events.
We change asynchronous logging into synchronous logging 
    and the overhead is acceptable
    as it only occurs when the follower receives \texttt{\small COMMIT} before handling the \texttt{\small NEWLEADER} message
    (synchronous logging is already used in v3.9.2).
The new implementation conforms to the new protocol specification.
We updated \mixedSpec{}s in Table~\ref{table:compo-spec}
    which passed model checking.

\section{Discussion} 
\label{sec:discussion}

Writing \tla{} specifications for distributed protocols and systems 
    has become a common practice~\cite{newcombe2015how_amazon,hackett2023understanding,amazon-use-tlaplus,azure-use-tlaplus,mongodb-use-tlaplus,linkedin-use-tlaplus,oracle-use-tlaplus}.
As articulated in~\cite{newcombe2015how_amazon},
    writing specifications enforces clear thinking, precise designs, and
    unambiguous documents.
We started from the protocol and system specifications (\S\ref{sec:existing_spec}).
However, it is clear to us that code-level implementation evolves fast and 
    inevitably deviates from the protocol and system designs due to
    performance optimizations.
As a consequence, deep bugs often reside in the model-code gaps (\S\ref{sec:background:gaps}).
We advocate for reconciling formal specifications with code implementations
    by fine-grained modeling of several important implementation aspects, such as 
    multithreading and non-atomic updates. 
We show that such modeling is beneficial and helps understand and address 
    a few complex, long-lasting issues in ZooKeeper.

We cope with the enlarged state space introduced by fine-grained modeling
    with mixed-grained specifications which are composed of 
    multi-grained specifications.
In essence, mixed-grained model checking is a divide-and-conquer strategy
    which leverages good modularity and loose coupling of distributed systems like ZooKeeper.
Apart from model checking, multi-grained specifications also bring benefits 
    such as enforcing precise thinking, communication, and documentation across 
    the ladder of abstractions (\S\ref{sec:verify-fix}).

Compared with implementation-level model checking~\cite{yang2009modist, guo2011practical, leesatapornwongsa2014samc, lukman2019flymc}, 
    our philosophy is different.
We intend to do checking at the model level,
    which is more efficient and scalable than 
    running heavy distributed system code. 
It is harder to reduce unnecessary overheads at the code level
    such as network and disk operations and code that is not targeted for verification.
Prior work~\cite{guo2011practical} developed dynamic interface reduction, but can only
    reduce to {\it node} locally.
Our goal is to develop models that effectively reflect the code
    and have the flexibility to choose granularities by {\it phases}.
\revision{
In our experience, it is not easy to define a clean local/non-local boundary for nodes
    in existing \tla{} specifications 
    due to the common use of global variables in \tla{}.}

Amazon recently shared their practice of using property-based testing with randomly generated
test input to exercise both the implementation and the executable model and checks whether they
agree~\cite{bornholt2021lightweight}.
We cannot directly apply this approach: (1) our model in \tla{} is not executable and (2) existing property-based
testing framework~\cite{proptest} cannot directly control event interleaving \textit{inside} the
system under test.
So, our conformance checker \revision{follows a top-down approach, i.e., using} the TLC model checker to generate event traces
and replays them in the implementation by deterministically controlling event interleaving\revision{, as used in SandTable~\cite{tang2024sandtable}.}
\revision{
Conformance can also be checked with a bottom-up approach, i.e.,
    generating implementation-level traces 
    and checking whether they are allowed by the model, 
    as adopted by VYRD~\cite{elmas2005vyrd}, CCF~\cite{cirstea2024validating} and etcd~\cite{etcd-tla}.
We choose the top-down approach as it can be easily reused 
    for deterministic replay and bug confirmation in the implementation 
    once some safety violation is found at the model level.}

We believe that our practice of writing multi-grained specifications is viable 
    and can be generalized beyond ZooKeeper.
As we show in \S\ref{sec:effort},
    the efforts on writing multi-grained specifications and instrumentations
    are reasonable and can be amortized, and the specifications are written by one person
    who knows deeply about 
    the protocol and the implementation.
On the other hand, 
\revision{
maintaining multi-grained models can be more expensive.
We have no silver bullet beyond running continuous conformance 
    checking upon code changes and updating the specifications accordingly.
Modern continuous integration 
    is an opportunity to update models incrementally.
Since most code changes are local, most modules and high-level specifications stay unchanged.
The cost of running conformance checking upon changes and updating specifications accordingly
    mainly lies in the changed modules and fine-grained specifications.
The mix-grained models can make continuous verification more efficient.
}
\section{Related Work} 
\label{sec:rw}

\revision{We documented our preliminary work in~\cite{ouyang2023zkspec}.
At that time, we were writing three types of specifications: (1) protocol specification, 
    (2) system specification as super-doc, and (3) test specification for testing 
    of ZooKeeper.
Those specifications were not coherent and could not meet the goal of verifying ZooKeeper implementation in this paper.

In~\cite{ouyang2023zkspec}, we defined a test specification as a refinement of a system specification. 
However, we later find that refinement is not the right approach \revisionred{to addressing model-code gaps (\S\ref{sec:background:gaps})};
    thus we no longer enforce refinement relations. 
Instead, we write fine-grained specifications to close model-code gaps and write coarse-grained specifications 
    to speed-up exploration of global states. 
For the same reason, unlike in~\cite{ouyang2023zkspec} where we viewed a system specification as an abstraction of a test specification, 
    we no longer enforce the abstraction relation during coarsening but only ensuring interaction preserving. 
For example, when specifying atomicity-related behavior, the coarse-grained specification 
    does not abstract away any variable from the fine-grained one, but has fewer transitions compared to the fine-grained one, 
    so no obvious abstraction or refinement relation exists.}

\para{Verification.}
\tla{} has been widely used for modeling and verifying distributed systems. 
Recent work~\cite{newcombe2015how_amazon,hackett2023understanding}
uses \tla{} to model and verify system designs, but not implementations.
Recently, \tla{} based techniques have also been developed to test and verify 
    distributed system implementations~\cite{tang2024sandtable, wang2023mocket, ouyang2023zkspec}.
From the tooling perspective, SandTable~\cite{tang2024sandtable} is a close approach.
Like \tool{}, SandTable models distributed systems in \tla{}, verifies systems using model checking
and ensures the specification quality using conformance checking.
Therefore, we believe that SandTable can effectively benefit from multi-grained specifications.
Notably, we notice that unlike \tool{}, SandTable cannot check thread interleaving
    because it intercepts at the system-call level and thus is hard to
    differentiate between user-level threads. 
This is one reason we choose to instrument application code so as to
    control user-level thread interleaving.

Programming languages with built-in model checking support~\cite{killian2007mace, killian07macemc, yabandeh2009crystalball, guerraoui2011lmc, hackett2023pgo, desai2013p}
can help build clean-slate verified distributed systems.
Unfortunately, it is hard for us to use them for ZooKeeper, which requires major revisions.
We were mostly looking for ``lightweight'' formal methods~\cite{bornholt2021lightweight}.

Compared to
implementation-level model checking~\cite{yang2009modist, simsa2010dbug, guo2011practical, leesatapornwongsa2014samc, lukman2019flymc},
we take a different approach by exploring state space at the \textit{model} level
to avoid code-level overhead,
    as discussed in \S\ref{sec:discussion}.
Among implementation-level model checkers, DeMeter~\cite{guo2011practical} also takes a divide-and-conquer
strategy and decomposes the problem of model checking a distributed system into model checking each {\it node} locally.
Differently, our approach decomposes the model checking problem 
    into model checking each {\it phase} and thus is complementary.

Prior work also explored model checking support for distributed systems~\cite{jfp, lauterburg09basset, saissi2013efficient}.
In particular, DBSS~\cite{saissi2013efficient} decomposes model checking by focusing on variables relevant to the property
being checked, and verifies protocol-level correctness.
Our decomposition is agnostic of any specific property,
and we focus on verifying whether the system correctly implements the protocol.

\revision{
Refinement checking and trace validation are also studied~\cite{tasiran2003using,elmas2005vyrd,cirstea2024validating,etcd-tla}.
Tasiran et al.~\cite{tasiran2003using} validates hardware designs by connecting the specification and simulation 
    and using model checking to monitor correctness and coverage. 
VYRD~\cite{elmas2005vyrd} implements I/O and view refinement checking to detect runtime refinement violations for concurrent programs.
Our approach verifies the implementation by exploring specification-level states, with the checking of specification-implementation conformance. 
The conformance checking shares a similar iterative process of selecting commit points in implementation and debugging the mapping between specification and implementation in \cite{tasiran2003using,elmas2005vyrd}.
}

Besides, deductive verification approaches
have been used to build verified distributed system
implementations~\cite{hawblitzel2015ironfleet, wilcox2015verdi, sergey2018disel, honore2021ado, sharma2023grove, sun2024anvil}.
Deductive verification does not need
a checker to explore state space.
However, deductive verification requires hard efforts
to write proofs and cannot be directly applied to existing distributed
systems.

\para{Bug finding.}
Many testing techniques~\cite{jepsen, tseitlin2013antifragile, gu2023acto, basiri2016chaos, alvaro2015ldfi_molly, sun2022sieve, chen2023rainmaker, wu2024legolas, kim2022modulo, lu2019crashtuner, chen2020cofi, ganesan2017redundancy, gunawi2011fate}
detect bugs in distributed systems by fault injection.
These tools typically inject faults randomly~\cite{tseitlin2013antifragile, ozkan2018randomized}
or focus on a system's vulnerable points
with manual guidance~\cite{kim2022modulo, jepsen, gunawi2011fate, chen2023rainmaker} or automated analysis~\cite{alvaro2015ldfi_molly, wu2024legolas, lu2019crashtuner, chen2020cofi}.
Besides, several projects~\cite{liu2017dcatch, liu2018fcatch, yuan2020effective} detect distributed concurrency bugs
caused by unexpected interleaving among node events by analyzing happen-before relationships
between events~\cite{liu2017dcatch, liu2018fcatch} or manipulating event ordering~\cite{yuan2020effective}.
Recent works have also used model checkers to generate test cases for distributed
systems~\cite{wang2023mocket, davis2020extreme_modelling}.
Developers at Amazon have applied property-based testing to test a production system
against executable specifications~\cite{bornholt2021lightweight}.
Despite their effectiveness in detecting bugs,
none of these techniques can verify a distributed system by exhaustively exploring its state space.

\section{Concluding Remarks} 
\label{sec:conclusion}

In this paper, we show that formal methods like \tla{} 
    can not only verify protocol and system designs,
    but also help verify system implementation
    by modeling important code-level behavior with conformance checking.
We advocate for the practice of multi-grained specification and show that 
    the composed mixed-grained 
    specifications provide useful capabilities to manage the state space,
    making model checking and verification efficient and more usable.
With formal methods like \tla{} being widely accepted 
    and adopted,
    we hope that our work leads to a forward step in empowering formal methods 
    to benefit distributed systems in practice.

\section*{Acknowledgment}
\revision{
We thank the anonymous reviewers and our shepherd, Serdar Tasiran, 
    for their insightful comments.
Huang's group is supported by the National Natural Science Foundation of China (62025202, 62372222), 
    the CCF-Huawei Populus Grove Fund (CCF-HuaweiFM202304),
    the Cooperation Fund of Huawei-NJU Next Generation Programming Innovation Lab (YBN2019105178SW38), and
    the Postgraduate Research \& Practice Innovation Program of Jiangsu Province (KYCX24\_ 0235).
Xu's group is supported in part by NSF CNS-2130560, CNS-2145295, and a
VMware Research Gift.
}

\appendix
\section{Descriptions of Detected Bugs}
\label{sec:bugs}

\revision{

We provide more information of the bugs in Table~\ref{table:bugs-found}. 
The detection of these bugs is described in \S\ref{subsec:effectiveness}.
Two of these bugs are known bugs,
    while the others are new bugs detected 
    during the process of verifying ZooKeeper using \tool{}.
\revisionred{All the bugs are deep safety bugs that are hard to trigger without model checking---each
    of them takes 
    tens of actions and tens of thousands of states to manifest (see Table~\ref{table:bugs-found}).}

\para{ZK-3023~\cite{ZK-3023}.}
The follower fails to catch up with the up-to-date committed data after data recovery is finished. 
It was caused by the asynchronous commit of the transactions during the Synchronization phase.
The bug is triggered when the leader handles \texttt{\small ACK} of \texttt{\small UPTODATE} before the follower commits the pending requests.
The bug was known but \tool{} still detected 
    it in the latest version by then. 
This bug was originally reported by a test, but the test cannot reliably trigger the buggy interleaving.
Our tool deterministically reproduced this bug by detecting violations of I-11 (bad states).

\para{ZK-4394~\cite{ZK-4394}.}
This bug could unexpectedly terminate data recovery,
    which can occur repeatedly and make the follower
    unavailable. 
When a follower cannot match the \texttt{\small COMMIT} message 
    to a received request in the Synchronization phase,
    it throws \texttt{\small NullPointerException}.
The bug is triggered when the follower, 
    after processing the \texttt{\small NEWLEADER} message,
    receives a \texttt{\small COMMIT} message before the \texttt{\small UPTODATE} message.
It was a known bug 
    but still in the latest version of ZooKeeper we verified. 

\para{ZK-4643~\cite{ZK-4643}.}
This bug results in data loss.
The implementation fails to
    guarantee that the follower atomically updates the history
    with the epoch (an atomic action in the original Zab protocol).
It manifests when a follower crashes after updating its epoch,
    becomes the new leader with stale committed history,
    and then truncates the committed data of others.
The triggering involves a follower crash
        between the update of epoch and the update of history, together with 
        two crashes of other nodes across three rounds of Election, Discovery, and Synchronization.
This is a new bug we detected.

\para{ZK-4646~\cite{ZK-4646}.}
This bug causes data loss.
The root cause lies in the asynchronous logging of the followers when the leader starts serving clients,
    which is assumed to be synchronously done by the protocol.
It manifests when the uncommitted data is seen by clients and then be removed later.
The bug is triggered when leader and followers all crash
    after a client reads the data that a majority of followers have not persisted in disk,
    and then one of the follower is elected as the new leader.
This is a new bug we detected. 

\para{ZK-4685~\cite{ZK-4685}.}
This bug fails data recovery among nodes,
    which increases recovery time and reduces system availability.
The root cause is that the leader fails to recognize an \texttt{\small ACK}, 
    which blocks the leader and then leads to the shutdown of all nodes from the Synchronization phase.
To trigger this, the follower replies to the leader with \texttt{\small ACK} of \texttt{\small PROPOSAL}
    before \texttt{\small ACK} of \texttt{\small NEWLEADER}
    when the leader is collecting a quorum of \texttt{\small ACK}s of \texttt{\small NEWLEADER}.
This is a new bug we detected.

\para{ZK-4712~\cite{ZK-4712}.}
This bug causes data inconsistency, as a follower keeps extra transactions 
in its log even after data recovery, making clients obtain inconsistent views from different servers.
It was caused by the asynchronous logging during the follower's shutdown.
To trigger it, a follower goes back to the Election phase with a non-empty request queue for logging,
and then updates its latest transaction ID before processing requests in the queue.
It is a new bug we detected. 

}

\newcommand{\cu}[1]{\widetilde{#1}}
\newcommand{\mc}[1]{\mathcal{#1}}
\newcommand{\set}[1]{\left\{#1\right\}}

\section{Proof Sketch of the Interaction Preservation Theorem}
\label{sec:appendix}

We prove that interaction-preserving coarsening does not affect the correctness of model checking.

We denote a specification $S$ that consists of $n$ modules as $S = \bigcup_{1 \le i \le n}M_{i}$,
    and we define $\widetilde{M_i}$ as a module obtained by coarsening $M_i$ 
    following the constraints of interaction preservation (\S\ref{appen:itr-prs}). 
\revision{
    $S_i$ is denoted as the specification by coarsening every other module except $M_i$, i.e.,
    $S_i = (\bigcup_{j \neq i}\widetilde{M_j}) \cup M_i$.}

Let the traces allowed by $S$ and $S_i$ be $T_S$ and $T_{S_i}$ respectively.
When we are only concerned with the states of the target module $M_i$, 
all traces in $T_S$ and $T_{S_i}$ are projected to $M_i$, 
which are denoted as $T_S|_{M_i}$ and $T_{S_i}|_{M_i}$.
Then we can talk about the equivalence relation between traces with respect to a target module,
which is defined as: $T_S \stackrel{M_i}{\sim} T_{S_i} \xlongequal{def} T_S|_{M_i} = T_{S_i}|_{M_i}$.

The safety of the coarsening is captured by the equivalence between traces, as in the following theorem:
\begin{namedtheorem}[Interaction Preservation]

Given $S = \bigcup_{1 \le i \le n}M_{i}$ and $S_i = (\bigcup_{j \neq i}\widetilde{M_j}) \cup M_i$,
we have $T_S \stackrel{M_i}{\sim} T_{S_i}$.

\end{namedtheorem}

\noindent \textit{Proof.} \ 
The basic idea of the proof is that, 
if the target module $M_i$ cannot distinguish 
whether it is interacting with the original module $M_j$ or the coarsened module $\cu{M_j}$, 
then the behavior of $M_i$ is not affected by the coarsening.

We define the notations used in the theorem and the proof
and present the rule for ensuring interaction preservation.
In the proof, we first present the condensation of traces ($T_S|_{M_i}$), to restrict our attention to the target module $M_i$.
Then we establish the equivalence between $T_S|_{M_i}$ and $T_{S_i}|_{M_i}$.

\subsection{Notations}

\para{\tla{} basics.}
In the \tla{} specification language, a system is specified 
    as a state machine by describing the possible initial states 
    and the allowed state transitions called $Next$. 
Specifically, the specification of system design contains 
    a set of \textit{system variables} $\mc{V}$. 
A \textit{state} is an assignment to the system variables. 
$Next$ is the disjunction of a set of actions $a_1\lor a_2\lor \cdots \lor a_p$, 
    where an \textit{action} is a conjunction of several clauses $c_1\wedge c_2\wedge \cdots \wedge c_q$.  
A \textit{clause} is either an \textit{enabling condition},
    or a \textit{next-state update}. 
An enabling condition is a state predicate 
    which describes the constraints the current state must satisfy, 
    while the next-state update describes 
    how variables can change in a step (i.e., successive states).

Whenever every enabling condition $\phi_a$ of an action $a$ 
    is satisfied in a given ``current'' state, 
    the system can transfer to the ``next'' state 
    by executing $a$, assigning to each variable the value specified by $a$. 
We use ``$s_1\stackrel{a}{\rightarrow}s_2$'' to denote that 
    the system state goes from $s_1$ to $s_2$ by executing action $a$, 
    and $a$ can be omitted if it is obvious from the context.
Such execution keeps going 
    and the sequence of system states forms a trace of system behavior.

%
A system usually consists of several modules, each implementing some specific function. 
For the \tla{} specification of a distributed system, we define: 

\setcounter{theorem}{0}

\begin{defi}[module]
    A module is a set of actions. All the modules form a partition of all actions in the specification.
\end{defi}

Assume that we write a specification 
    $S = \bigcup_{1 \le i \le n}M_{i}$ for the system under verification.
Our target module is $M_i$, 
    and we coarsen every other module, obtaining
    $S_i = (\bigcup_{j \neq i}\widetilde{M_j}) \cup M_i$.
The coarsening ensures interaction preservation (\S\ref{appen:itr-prs}).

We define the set of all possible traces allowed by $S$ (resp. $S_i$) as $T_S$ (resp. $T_{S_i}$).
$T_S$ and $T_{S_i}$ are different, 
    and $T_S$ usually has a much larger size than $T_{S_i}$.
With target module $S_i$ in mind, 
    we omit unrelated details in the traces by \textit{condensation} (\S\ref{appen:conden}), 
    and then show that the two sets of traces are equivalent with respect to the target module $M_i$ (\S\ref{appen:equiv}).

\subsection{Interaction Preservation} 
\label{appen:itr-prs}

Modules interact with each other through the system variables.
To capture this, we first define the \textit{dependency variable} of an action and that of a module:
\begin{definition}[dependency variable] \label{Def: DepVar}
    Suppose module $M=\{a_1,a_2,\cdots ,a_m\}$, dependency variables of $M$, denoted as $\mathcal{D}_{M}$, is obtained recursively according to the following rules:
    \begin{enumerate}[leftmargin=*]
        \item For any action $a_i\in M$, its dependency variables $\mathcal{D}_{a_i}$ are the variables which appear in some enabling condition $\phi_{a_i}$ of $a_i$. 
        \item $\bigcup_{1\leq i\leq m}\mathcal{D}_{a_i}\subseteq \mathcal{D}_M$. That is, the dependency variables of each action in $M$ belong to $\mathcal{D}_M$.
        \item For any $v \in \mathcal{D}_{M}$ and any action $a_i\in M$, if the next-state update of $a_i$ assigns to $v$ a value calculated from multiple variables (denoted by variable set $V_{dep}$), then $V_{dep} \subseteq \mc{D}_M$. This is due to transitivity of the dependency relation, i.e., if $M$ depends on some variable $v$ and $v$ depends on another variable $w$, then $M$ also depends on $w$.
    \end{enumerate}
\end{definition}

\noindent Given the definitions above, we can now say that module $M_j$ interacts with $M_i$ by modifying $\mc{D}_{M_i}$.

The notion of dependency variable alone is not sufficient to capture interactions among modules, since even if $D_{M_i}$ are not modified by some action in $M_j$, $M_i$ may still be affected indirectly.
Suppose $x \in \mathcal{D}_{M_i}$, 
an action in another module $M_j$ assigns to $x$ the value of $y$ (note that $y$ will not be added to $\mc{D}_{M_i}$ by the Rule 3 in Definition \ref{Def: DepVar}, since $x$ is assigned the value of $y$ in module $M_j$, not in $M_i$). 
In this case, any assignment to $y$ may also change the value of $x$ in subsequent actions. 
To capture such indirect interactions among modules, we define the set of interaction variables $\mc{I}$:
\begin{definition}[interaction variable]
    Suppose the specification contains $k$ modules: $M_1,\cdots,M_k$. 
    The set of interaction variables $\mathcal{I}$ is calculated recursively according to the following rules:
    \begin{enumerate}[leftmargin=*]
        \item $\bigcup_{1\leq i<j\leq k}(\mathcal{D}_{M_i}\cap \mathcal{D}_{M_j}) \subseteq \mathcal{I}$. That is, if a variable is a dependency variable of multiple modules, then it belongs to $\mathcal{I}$.
        
        \item For any $v \in \mathcal{I}$ and any module $M_i$, if an action $a \in M_i$ assigns to $v$ a value calculated from multiple variables (denoted by set $V_{intr}$), then add all variables in $V_{intr}\setminus \mathcal{D}_{M_i}$ to $\mathcal{I}$. That is, the value assigned to an interaction variable by any action in $M_i$ should be calculated from values of variables in interaction variables or dependency variables of the module, i.e., $\mathcal{I} \cup \mathcal{D}_{M_i}$.
        
        \item For any variable $v \in \mathcal{D}_{M_i}\setminus \mathcal{I}$ in any module $M_i$, if an action assigns to $v$ a value calculated from multiple variables (denoted by set $V'_{intr}$), then add all variables in $V'_{intr}\setminus \mathcal{D}_{M_i}$ to $\mathcal{I}$. That is, the value assigned to a ``internal'' variable of $M_i$ by any action should be calculated from values of interaction variables or from values of dependency variables of the module, i.e., $\mathcal{I}\cup \mathcal{D}_{M_i}$.
    \end{enumerate}
\end{definition}

Note that Rule 1 of this definition is conservative. 
Some variable $x$ in both $\mc{D}_{M_i}$ and $\mc{D}_{M_j}$ 
may not convey any interaction between $M_i$ and $M_j$. 
However, in practice such cases are rare (see the case study in \S\ref{sec:lfm}).

The coarsening preserves interaction if: 
    (1) all dependency variables of the target module,
    as well as all interaction variables remain unchanged after the coarsening;
    (2) all the updates of the dependency variables and interaction variables remain unchanged after the coarsening.
Put it differently, 
    only variables in $\mc{V} \backslash (\mc{I} \cup \mc{D}_{M_i})$ 
    and state updates only involving such variables 
    can be omitted during coarsening.

We use the interaction preservation rules to write coarse-grained specifications, 
    while not affecting state-space exploration of the target module.
The mixed-grained specifications 
    help us tame state-space explosion.

\subsection{Condensation} \label{appen:conden}

The basic idea of condensation is 
    to merge a set of equivalent states into one state.
The equivalence between states is established 
    based on the projection to the target module $M_i$.
After condensation of equivalent states in one trace, 
    we can also condense equivalent traces, also based on the projection to $M_i$.
Only after condensation, can we define the equivalence between two sets of traces in \S\ref{appen:equiv}.

For specifications $S = \bigcup_{1 \le i \le n}M_{i}$, 
    the state $s$ is defined as the valuation of all variables in $S$.
State $s|_{M_i}$ is defined as the projection of $s$ to $M_i$, 
    i.e., $s|_{M_i}$ is the valuation of all variables in $\mc{D}_{M_i} \cup \mc{I}$.

For state $s$ allowed by specification $S$ 
    and state $\cu{s}$ allowed by specification $S_i$, 
    we say that $s$ is equivalent to $\cu{s}$ 
    if $s|_{M_i} = \cu{s}|_{M_i}$.
This equivalence relation 
    is denoted as $s\stackrel{M_i}{\sim} \cu{s}$.
Similarly, we can also define 
    the equivalence between two states $s_1$ and $s_2$ 
    both from the same trace. 
We have $s_1 \stackrel{M_i}{\sim} s_2$ 
    if $s_1|_{M_i} = s_2|_{M_i}$.
Basically, the equivalence in both cases means that 
    all variables in $\mc{D}_{M_i}\cup \mc{I}$ remain unchanged in two states.

For a state transition $s_1\rightarrow s_2$, 
    the transition is \textit{interesting} to $M_i$, if $\neg (s_1 \stackrel{M_i}{\sim} s_2)$.
The transition is \textit{not-interesting}, if $s_1 \stackrel{M_i}{\sim} s_2$.
In other words, in an interesting transition, 
    one or more variables in $\mc{D}_{M_i}\cup\mc{I}$ are updated. 
In a not-interesting transition, 
    all variables in $\mc{D}_{M_i}\cup \mc{I}$ remain unchanged.

We define the \textit{condensation} of a trace as omitting not-interesting transitions. 
Given a trace $t$ allowed by $S$ or $S_i$, 
    for any transition $s_1\rightarrow s_2$ in $t$, 
    if the transition is not-interesting, we condense the transition 
    by merge $s_1$ and $s_2$ to a set of equivalent states $\set{s_1, s_2}$.
After the condensation, a trace is 
    the transitions from one set of equivalent states to another.
Since all states merged are equivalent, 
    they can be viewed as just one state, 
    and the condensed trace can still be viewed as state transitions. 

After the condensation of each trace, 
    we can also condense a set of traces.
For any two condensed traces allowed by some specification $S$, 
    if the two states of the same index are equivalent for every index, 
    we deem these two traces equivalent,
    and they are merged into a set of equivalent traces.
Similarly, since all traces merged are equivalent,
    they can be viewed as just one trace.

\subsection{Equivalence} \label{appen:equiv}

Given that every trace is condensed 
    and the set of traces is condensed, 
    we can construct the equivalence relation between two sets of traces $T_S$ and $T_{S_i}$.

After the condensation, any state transition is some update of state $s|_{M_i}$.
Given the equivalence relation between states, 
    the equivalence relation can also be established between two sets of traces $T_S$ and $T_{\cu{S}}$.
In both cases, the equivalence relations are with respect to $M_i$.

We define: $T_S \stackrel{M_i}{\sim} T_{S_i} \xlongequal{def} T_S|_{M_i} = T_{S_i}|_{M_i}$.
Here, by $T_S|_{M_i} = T_{S_i}|_{M_i}$, we mean that: 
    (1) there is a bijective mapping between $t\in T_S$ and $\cu{t}\in T_{S_i}$;
    (2) for every state $s_k\in t$ and $\cu{s}_k\in \cu{t}$, $s_k\stackrel{M_i}{\sim} \cu{s}_k$.

The bijective mapping between $t$ and $\cu{t}$ can be derived from the fact that the coarsening ensures interaction preservation.
The mapping is constructed based on the induction on state index $k$ in the trace.

The initial state of model checking is set the same (with projection to $M_i$) before and after the coarsening.
Thus for any trace $t\in T_S$ and $\cu{t}\in T_{S_i}$, $s_0 \stackrel{M_i}{\sim} \cu{s}_0$.

By the induction hypothesis, 
    for state $s_k\in t$, we have exactly one corresponding state $\cu{s}_k$ 
    with $s_k \stackrel{M_i}{\sim} \cu{s}_k$.
Now consider the transition $s_k\rightarrow s_{k+1}$.
After the trace condensation, $s_k$ and $s_{k+1}$ are not equivalent, 
    and $s_k\rightarrow s_{k+1}$ must involve 
    update of one or more variables in $\mc{D}_{M_i}\cup \mc{I}$.

According to interaction preservation rule (\S\ref{appen:itr-prs}), the variables in $\mc{D}_{M_i}\cup \mc{I}$ and updates of these variables in any action remain unchanged after the coarsening.
So we have the same updates on the same set of variables, 
    i.e. variables in $\mc{D}_{M_i}\cup \mc{I}$, for $s_k$ and $\cu{s}_k$.

Note that each state transition is deterministic.
This is ensured by the fact that 
    in all our specifications $S$ and $S_i$, 
    any state updates in any action is deterministic.
The same updates on the same set of variables 
    deterministically produce one unique next state,
    given that we only consider the projection to $M_i$. 
That is, we are only concerned of state updates of the target module $M_i$ 
    and ignore other information in the states and the state transitions.
Thus, $s_k$ and $\cu{s}_k$ both have their unique successors $s_{k+1}$ and $\cu{s}_{k+1}$ respectively, and $ s_{k+1} \stackrel{M_i}{\sim} \cu{s}_{k+1}$.

By induction, we have the desired mapping between traces in $T_S|_{M_i}$ and $T_{S_i}|_{M_i}$.
This gives us that $T_S \stackrel{M_i}{\sim} T_{S_i}$, which ensures the safety of model checking after the coarsening. 
\qed

\balance

\bibliographystyle{acm}
\bibliography{main}

\end{document}